\documentclass[aps,twocolumn,prd,nofootinbib,longbibliography]{revtex4-2}

\usepackage{amsmath, amssymb, amsfonts, amsthm, latexsym, epsfig, mathrsfs, xcolor, bbm, slashed,braket}

\usepackage[inline]{enumitem}

\usepackage{setspace}
\usepackage[marginal, multiple]{footmisc}

\usepackage[T1]{fontenc}
\usepackage[utf8]{inputenc}
\usepackage{lmodern}
\usepackage[normalem]{ulem}

\usepackage[unicode, colorlinks, allcolors=blue!70!black, linktocpage, pdfusetitle]{hyperref}
\usepackage[all]{hypcap}

\usepackage{tikz}
\usetikzlibrary{arrows.meta}

\numberwithin{equation}{section}

\usepackage{cleveref}

\usepackage{cancel}

\usepackage{microtype}

\setlist[enumerate]{noitemsep, label=(\arabic*), ref=(\arabic*)}

\newlist{condlist}{enumerate}{2}
\setlist[condlist,1]{noitemsep, label=(\arabic*), ref=(\arabic*)}
\setlist[condlist,2]{noitemsep, label=(\alph*), ref=(\arabic{condlisti}.\alph*)}
\crefname{condlisti}{condition}{conditions}
\crefname{condlistii}{condition}{conditions}

\renewcommand\thesection{\arabic{section}}
\renewcommand\thesubsection{\arabic{subsection}}

\makeatletter
\def\p@subsection{\thesection.}
\def\p@subsubsection{\thesection.\thesubsection.}
\makeatother

\theoremstyle{plain}

\theoremstyle{definition}

\theoremstyle{remark}

\crefname{equation}{Eq.}{Eqs.}
\Crefname{equation}{Equation}{Equations}

\crefname{section}{Sec.}{Secs.}
\crefname{appendix}{Appendix}{Appendices}
\crefname{figure}{Fig.}{Figs.}

\crefname{definition}{Def.}{Defs.}
\crefname{prop}{Prop.}{Props.}
\crefname{lemma}{Lemma}{Lemmas}
\crefname{corollary}{Cor.}{Cors.}
\crefname{thm}{Theorem}{Theorems}
\crefname{remark}{Remark}{Remarks}

\crefname{ass}{Assumptions}{Assumptions}
\crefname{property}{Properties}{Properties}

\newcommand{\be}{\begin{equation}}
\newcommand{\ee}{\end{equation}}

\newcommand{\mc}{\mathcal}

\newcommand{\ms}{\mathscr}

\newcommand{\Lie}{\pounds}
\newcommand{\hatLie}{\Lie\kern-0.25em\hat{\vphantom{\Lie{}}}\kern0.25em}

\newcommand{\scri}{\ms I}

\begin{document}

\title{Gravitationally Mediated Entanglement: Newtonian Field vs. Gravitons}

\author{Daine L. Danielson}\email{daine@uchicago.edu}
\author{Gautam Satishchandran}\email{gautamsatish@uchicago.edu}
\author{Robert M. Wald}\email{rmwa@uchicago.edu}
\affiliation{Enrico Fermi Institute and Department of Physics, The University of Chicago, 933 East 56th Street, Chicago, Illinois 60637, USA}

\date{\today}

\begin{abstract}
\noindent We argue that if the Newtonian gravitational field of a body can mediate entanglement with another body, then it should also be possible for the body producing the Newtonian field to entangle directly with on-shell gravitons. Our arguments are made by revisiting a gedankenexperiment previously analyzed by Belenchia \textit{et al.}~\cite{Wald2018,Belenchia:2019gcc}, which showed that a quantum superposition of a massive body requires both quantized gravitational radiation and local vacuum fluctuations of the spacetime metric in order to avoid contradictions with complementarity and causality. We provide a precise and rigorous description of the entanglement and decoherence effects occurring in this gedankenexperiment, thereby significantly improving upon the back-of-the-envelope estimates given by Belenchia \textit{et al.} and also showing that their conclusions are valid in much more general circumstances. As a by-product of our analysis, we show that under the protocols of the gedankenexperiment, there is no clear distinction between entanglement mediated by the Newtonian gravitational field of a body and entanglement mediated by on-shell gravitons emitted by the body. This suggests that Newtonian entanglement implies the existence of graviton entanglement and supports the view that the experimental discovery of Newtonian entanglement may be viewed as implying the existence of the graviton.

\end{abstract}
\maketitle

\section{Introduction}
General relativity and quantum field theory are the two fundamental pillars of modern physics. Their union in the form of a theory of quantum gravity remains the most significant open issue in theoretical physics. Although one can formulate an essentially satisfactory theory of linearized quantum gravity perturbed off of some fixed background spacetime, severe difficulties arise in formulating a nonperturbative theory of quantum gravity. While strong arguments can be given that gravity should be quantized \cite{Bronstein,Page_1981,Eppley_1977,Mattingly_2006,Carlip_2008,Giampaolo_2019}, these difficulties have led some to suggest that gravity may be fundamentally classical, that the description of gravity with quantum mechanics requires a radical modification of quantization \cite{Hossenfelder:2010,Penrose_2014,DIOSI_1987}, or that the question of quantization is ill-posed \cite{DYSON_2013}. Of central importance to this debate is the prediction of quantized gravitational radiation in the form of gravitons, the existence of which has not yet been verified experimentally.

As already noted by Feynman in the 1950's \cite{Dewitt_2011,Zeh_2008}, some key issues regarding the quantization of gravity can be explored by considering the gravitational field sourced by a quantum superposition of a massive body. Due to recent advances in maintaining coherent spatial superpositions,\footnote{Spatial superpositions of masses on the scale of $10^{5}$ amu over distances of order microns have been achieved \cite{Gerlich_2011,Eibenberger_2013,Oriol_2017,Fein_2019} and recent proposals have suggested up to nanogram scale superpositions \cite{Pino_2018,brand_2017}.} many actual experiments involving such superpositions have recently been proposed \cite{FORD_1982,Lindner_2005,Bahrami_2015}. Given the rapid progress toward proposed ``low-energy'' tabletop experiments \cite{Bose_2017,Marletto_2017,Carney_2019, Haine_2021, Qvarfort_2020, Carlesso_2019, Howl_2021, Matsumura_2020, Pedernales_2021, Liu_2021, Datta_2021, Gonzalez-Ballestero:2021, Krisnanda:2020uh, doi:10.1126/sciadv.abg2879, CHRISTODOULOU201964, Bose:2022uxe}, it is of interest to understand what such low-energy phenomena might teach us about the fundamental nature of quantum gravity.

The analysis by Belenchia \textit{et al.}~\cite{Wald2018,Belenchia:2019gcc} of a gedankenexperiment originally proposed by \cite{Mari2009} provides strong evidence that low-energy experiments can probe quantum field theoretic aspects of gravity. In this gedankenexperiment, an experimenter, Alice, puts a massive body (hereinafter referred to as a ``particle'') into a quantum superposition at different spatial locations. At a later time, she recombines the particle and determines its quantum coherence. In the meantime---at a spacelike separation from the recombination portion of Alice's experiment---another experimenter, Bob, measures the Newtonian gravitational field of Alice's particle to try to determine its position. If Bob succeeds, then by complementarity, Alice's particle must be decohered. But, if Bob influences the state of Alice's particle, then causality would be violated. The analysis by Belenchia \textit{et al.}~\cite{Wald2018,Belenchia:2019gcc} showed that, in order to avoid contradictions with complementarity or causality, quantum gravity must have fundamental features of a quantum field theory at low energies, specifically the quantization of gravitational radiation (which decoheres Alice's particle without the presence of Bob) and local vacuum fluctuations (which limits Bob's ability to measure the position of Alice's particle). 

However, the analysis of~\cite{Wald2018,Belenchia:2019gcc} made only back-of-the-envelope estimates for the decoherence effects associated with Alice's recombination and Bob's measurement. Furthermore, it considered only a particular type of measurement that Bob might make. An important purpose of this paper is to reanalyze this gedankenexperiment, allowing Bob to make any measurement whatsoever in the region spacelike separated from Alice's recombination region. We give a precise analysis of the decoherence associated with radiation emitted by Alice's particle and the decoherence associated with Bob's measurement. We thereby confirm in a rigorous way the conclusions that had been drawn in~\cite{Wald2018,Belenchia:2019gcc} from their back-of-the-envelope estimates. 

Our analysis sheds additional light on the issue of whether tabletop experiments probe only quantum properties of the Newtonian gravitational field \cite{Anastopoulos_2018}. Since Bob sees only the Newtonian gravitational field of Alice's superposition during the time of his measurement, it is natural to view this Newtonian field as mediating entanglement between Bob and Alice. Indeed, if Alice decides to recombine her body at a much later time, the resulting correlations between the state of Bob's measuring apparatus and the state of Alice's particle must be viewed as having been mediated by the Newtonian field of Alice's particle. However, we will show that if Alice follows her protocol and recombines her particle in a region spacelike separated from Bob's measurements, then it is much more natural to view Bob as having measured on-shell gravitons that were emitted by Alice's particle; i.e., although Bob may believe that he is measuring a Newtonian gravitational field, he is actually measuring long wavelength gravitons. This viewpoint makes it clear that if the protocols of the gedankenexperiment are followed, then Bob is merely a ``bystander'' and his measurements have no relevance to the decoherence of Alice's particle.

Thus, in the circumstances of our gedankenexperiment, there is no clear distinction between entanglement of Alice's particle with Bob's apparatus that is mediated by a Newtonian field and entanglement of Alice's particle with gravitons that then interact with Bob's apparatus. This suggests that, in more general circumstances, entanglement mediated by a Newtonian field is not fully distinguishable from entanglement with gravitons and, hence, that the experimental discovery of entanglement by a Newtonian field may be viewed as evidence for existence of the graviton as a fundamental particle of nature.\footnote{Additional arguments for this conclusion have been given in \cite{Carney_2021}.} Furthermore, our analysis provides support for the conclusions of \cite{Belenchia:2019gcc} that the Newtonian field itself can store and transmit quantum information. 

In \cref{sec:review}, we review the gedankenexperiment of \cite{Mari2009} and its analysis by \cite{Wald2018}. In \cref{sec:decAliceBob} we analyze the decoherence effects associated with the emission of quantized radiation by Alice's particle and the decoherence effects associated with measurements made by Bob. In \cref{sec:gedankres} we reanalyze the gedankenexperiment in a more precise way and provide a proof that no violations of causality or complementarity occur. Some further remarks and conclusions are given in \cref{sec:summconc}. 

Throughout the paper, we will work in Planck units where $G=c=\hbar=1$.

\section{The gedankenexperiment of Mari \textit{et al.} and its resolution by Belenchia \textit{et al.}}
\label{sec:review}

In this section we review the gedankenexperiment initially proposed by Mari \textit{et al.} \cite{Mari2009} and its resolution given by Belenchia \textit{et al.} \cite{Wald2018}. There are electromagnetic and gravitational versions of this gedankenexperiment. For simplicity and definiteness, we shall first focus on the electromagnetic version and then discuss the modifications to the analysis needed for the gravitational case.

\begin{figure}[h!]
    \centering
    \includegraphics[width=\linewidth]{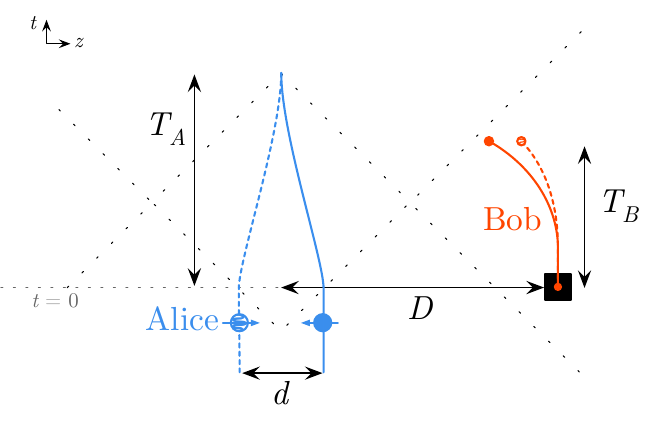}
     \caption{The setup for the gedankenexperiment of \cite{Mari2009}, as analyzed in \cite{Wald2018}. Alice's particle (in blue) is originally in the superposition state \cref{eq:Aliceinst} with the two wave packets separated by distance $d$. Bob is at a distance $D\gg d$ from Alice and, at a prearranged time, he releases a particle (in orange) from a trap and attempts to gain information about which path Alice's particle took by determining the strength of the Coulomb/Newtonian field of Alice's particle. Meanwhile, at a corresponding prearranged time, Alice recombines her particle and determines its coherence as described in the text. Bob does his measurement within time $T_{B} < D$ and Alice recombines her particle in time $T_{A} < D$, so their actions are performed in spacelike separated regions.}
      \label{fig:originexp}
\end{figure}

The gedankenexperiment is illustrated in \cref{fig:originexp}. At some time in the distant past, Alice sent a charged particle with spin in the positive $x$ direction through a Stern-Gerlach apparatus that is oriented in the $z$ direction. We assume that this process was done sufficiently slowly so as to produce negligible radiation and that Alice's particle can be described by ordinary, nonrelativistic quantum mechanics.
After going through the Stern-Gerlach apparatus, her particle is then in a superposition state of the form
\be 
\label{eq:Aliceinst}
\frac{1}{\sqrt{2}}\big(\ket{\uparrow;A_{1}}+\ket{\downarrow;A_{2}}\big)
\ee 
where $\ket{A_1}$ and $\ket{A_2}$ describe spatially separated wave packets and $\ket{\uparrow}$ and $\ket{\downarrow}$ represent eigenstates of $z$ spin. At a prearranged time, Bob attempts to determine which path Alice's particle followed by measuring the Coulomb field of Alice's particle. One way that Bob could do this is to release a charged particle from a trap at the prearranged time; if Alice's particle takes the right path in \cref{fig:originexp}, the Coulomb field near Bob will be stronger and the motion of Bob's particle will be influenced more, so by measuring the position of his particle at a later time, Bob can obtain some ``which-path'' information about Alice's particle. At a corresponding, prearranged time, Alice recombines her particle by putting it through a ``reversing Stern-Gerlach apparatus'' \cite{Mari2009,Bose_2017}. Alice then determines the coherence of her recombined particle by measuring its spin in the $x$ direction.\footnote{In the version of the gedankenexperiment discussed in \cite{Wald2018}, Alice determines the coherence of her particle by performing an interference experiment on the particle wave packets. An alternative resolution of that version of the gedankenexperiment was proposed in \cite{Rydving_2021}, based upon postulating fundamental limits to the ability to resolve interference fringes as originally proposed by \cite{Baym_2009}. This alternative resolution would not be applicable to the version of the gedankenexperiment being considered here.} If her had particle maintained perfect coherence, it would evolve back to an eigenstate of spin in the positive $x$ direction. By contrast, if the components of the original superposition \cref{eq:Aliceinst} had completely decohered, Alice would find that the spin is in the positive $x$ direction only $50\%$ of the time. By repeating the gedankenexperiment as many times as necessary, Alice can build up good statistics on the $x$ spin and thereby determine the degree of decoherence of her particle. By the prearranged protocol, the spacetime region in which Alice does the recombination and spin measurement is spacelike separated from the region in which Bob does his measurements, as illustrated in \cref{fig:originexp}.

This gedankenexperiment appears to lead to a contradiction with complementarity or causality. If Bob acquires any which-path information from his measurement, the state of Bob's particle must be correlated with Alice's to some degree. In that case, by complementarity, Alice's particle cannot be in a perfectly coherent superposition and she will find her particle to have spin in the negative $x$ direction some of the time. On the other hand, since Bob and Alice perform their actions in spacelike separated regions, by causality, it is impossible for Bob's measurements to have any effect on Alice's results, so the fact that he obtained some which-path information cannot degrade the coherence of Alice's particle. So, if Bob's measurement does not influence Alice's spin measurement, we would appear to have a violation of complementarity, whereas if Bob's measurement does influence Alice's spin measurement, we have a clear violation of causality. 

A resolution of this apparent paradox was given in \cite{Wald2018}. This resolution is based upon Bob's limitations in acquiring which-path information due to vacuum fluctuations and Alice's limitations in maintaining coherence due to the emission of entangling radiation. Bob's limitations due to vacuum fluctuations can be estimated as follows. In the electromagnetic case, the difference of the Coulomb electric fields associated with the different paths of Alice's particle is given by 
\be
E \sim \frac{\mc{D}_{A}}{D^{3}}
\ee
where $D$ is the distance between Alice and Bob and $ \mc{D}_{A} = q_A d$, where $q_A$ is the charge of Alice's particle and $d \ll D$ is the distance between the two paths of Alice's particle. If Bob must perform his measurement in time $T_B $, the difference in the final position of his particle due to the difference in the Coulomb fields of Alice's particle is
\be
\delta x \sim  \frac{q_{B}}{m_{B}} \frac{\mc{D}_{A}}{D^{3}} T^2_B 
\label{dxem}
\ee
where $q_B$ is the charge of Bob's particle and $m_B$ is its mass. On the other hand, vacuum fluctuations of the electromagnetic field produce fluctuations in the position of Bob's particle of order
\be
\Delta x \sim  \frac{q_{B}}{m_{B}}.
\label{Dxem}
\ee
Thus, on account of the ``noise'' due to vacuum fluctuations, Bob can acquire significant which-path information only if
\be 
\label{eq:whichpath}
\frac{\mc{D}_{A}}{D}>\bigg(\frac{D}{T_{B}}\bigg)^{2}.
\ee
In particular, if Bob abides by his protocol $T_B < D$, he can acquire significant which-path information only when $\mc{D}_{A} > D$.
 
Alice's limitations on maintaining coherence due to radiation can be estimated as follows. When Alice recombines her particle over a time $T_A$, she reduces the initial effective dipole $\mc{D}_{A}$ to zero. By the Larmor formula, this should result in emission of entangling radiation corresponding to an average energy flux $\sim (\mc{D}_{A}/T_A^2)^2$. Thus the total energy radiated should be $\sim \mc{D}^2_{A}/T_A^3$. This radiation should be composed of photons of frequency $\sim 1/T_A$. Thus the total number of entangling photons emitted when Alice recombines her particle should be
\be
N \sim \frac{\mc{D}^2_{A}}{T^2_A}.
\label{nem}
\ee
If $N > 1$, then Alice's particle will undergo significant decoherence due to entanglement with radiation, independent of what Bob does. In particular, if Alice abides by her protocol $T_A < D$, she can maintain coherence only when  $\mc{D}_{A} < D$.

The above estimates allow one to provide the resolution given in \cite{Wald2018}. If $\mc{D}_{A} > D$, then Bob can acquire significant which-path information, so by complementarity, Alice's particle must correspondingly be significantly decohered. However, in this case the radiation emitted when Alice does her recombination will decohere her particle independently of what Bob does, so there is no reason to believe that Bob's measurement ``caused'' the decoherence, i.e., there is no obvious violation of causality. On the other hand, if $\mc{D}_{A} < D$, then Alice should be able to largely maintain the coherence of her particle during the recombination. But in this case, Bob cannot acquire significant which-path information, so complementarity does not imply decoherence of Alice's particle and, again, there is no obvious violation of causality.

The analysis of the gravitational version of the gedankenexperiment within the context of linearized quantum gravity is very similar, with the main difference being the replacement of ``dipole'' by ``quadrupole.'' Alice's original separation of the particle into a superposition of two paths does not produce an effective dipole on account of conservation of center of mass---her laboratory must produce an equal and opposite compensating mass dipole. Thus, \cref{dxem} gets replaced by
\be
\delta x \sim  \frac{\mc{Q}_{A}}{D^{4}} T^2_B
\ee
where ${\mc Q}_A = m_A d^2$, where $m_A$ is the mass of Alice's particle. The replacement of \cref{Dxem} is the Planck length which, in our units, is given by 
\be
\Delta x \sim 1.
\ee
Since Alice now dominantly would emit quadrupole radiation during her recombination, the replacement of \cref{nem} is
\be
N \sim \frac{\mc{Q}^2_{A}}{T^4_A}.
\ee
Suppose that Bob and Alice follow their protocols, so that $T_B < D$ and $T_A < D$. Then if $\mc{Q}_{A} > D^2$, Bob can acquire significant which-path information but Alice decoheres her particle with gravitational radiation independent of what Bob does. Conversely, if $\mc{Q}_{A} < D^2$, then Alice should be able to largely maintain the coherence of her particle during the recombination, but Bob cannot acquire significant which-path information. Thus, as in the electromagnetic case, there is no obvious contradiction with complementarity or causality.

The above analysis of \cite{Wald2018} resolves the apparent paradox posed by the gedankenexperiment. Interestingly, this analysis shows that both quantized radiation and vacuum fluctuations are essential for resolving the paradox. Nevertheless, there are some unsatisfactory aspects of this analysis. In particular, only back-of-the-envelope estimates of the various effects were made, so only a rough, order of magnitude relation was obtained between the decoherence due to radiation during Alice's recombination and the decoherence associated with Bob's measurement. Furthermore, one might consider ways in which Bob might improve his ability to obtain which-path information. For example, suppose that Bob, together with $n-1$ assistants, sets up $n$ separate experiments like the one pictured in \cref{fig:originexp} to measure the Coulomb/Newtonian field of Alice's particle. Suppose that each of these $n$ experiments are done in regions that are spacelike separated from Alice's recombination region and spacelike separated from each other. If each of these experiments could be treated as independent, one would obtain an improvement of $1/\sqrt{n}$ in Bob's ability to overcome the noise due to vacuum fluctuations. Bob would then be able to obtain a corresponding improvement in his acquisition of which-path information, so if $n$ could be taken to be sufficiently large, we would again get a contradiction with complementarity or causality. In fact, vacuum fluctuations over spacelike separated regions are correlated, so it is not obvious that the $n$ experiments can be treated as independent. But it also is not obvious that a scheme of this sort would not work. Thus, while the analysis of \cite{Wald2018} is satisfactory for indicating that there are no obvious contradictions with complementarity or causality, it is not adequate for conclusively showing that no such contradictions can ever occur in this type of gedankenexperiment.

As already stated in the Introduction, an important purpose of this paper is to improve the analysis of \cite{Wald2018} by giving much more precise versions of the above estimates. We will thereby show in a much more rigorous way that no contradictions with complementarity or causality can occur in this type of gedankenexperiment. As a very important by-product, we will also obtain additional insights into how the state of Alice's particle and the state of Bob's apparatus become correlated. Should this correlation be viewed as being mediated by the Coulomb/Newtonian field of Alice's particle or by on-shell photons/gravitons emitted during the recombination process? We will show that both viewpoints are correct, i.e., they are equivalent descriptions of the same phenomena. We begin in the next section by giving precise descriptions of the decoherence due to Alice and the decoherence due to Bob.

\section{Decoherence due to Alice and Decoherence due to Bob}
\label{sec:decAliceBob}

In this section, we give a more precise characterization of the decoherence of Alice's particle due to radiation emitted when she recombines her particle and the decoherence associated with Bob's measurements. These characterizations will be used in the next section to reanalyze the gedankenexperiment. In this section we will explicitly discuss the electromagnetic version of the gedankenexperiment, since the language and concepts are more familiar in this context. However, exactly the same discussion applies to the gravitational case, with appropriate substitutions of ``graviton'' for ``photon,'' ``Newtonian'' for ``Coulomb,'' etc.

\subsection{Decoherence Due to Alice}
\label{subsec:decAlice}

We first consider the decoherence of Alice's particle that would occur in the absence of Bob or any other external influence.

Previously, we stated that after Alice sends her particle through a Stern-Gerlach apparatus at an early time, the particle is in the superposition state \cref{eq:Aliceinst}. However, this expression ignores the electromagnetic field, which is in a different state depending upon the state of Alice's particle. Heuristically, the state of the total system should be of the form
\be 
\label{eq:Aliceinst2}
\frac{1}{\sqrt{2}}\big(\ket{\uparrow;A_{1}} \otimes \ket{\psi_1}+\ket{\downarrow;A_{2}} \otimes \ket{\psi_2}\big)
\ee 
where states $\ket{\psi_{1}}$ and $\ket{\psi_{2}}$ formally correspond to coherent states of the Coulomb field of Alice's particle in states $\ket{\uparrow;A_1}$ and $\ket{\downarrow;A_2}$ respectively. However, this is only a formal expression because the ``Coulomb states'' $\ket{\psi_{1}}$, $\ket{\psi_{2}}$ are not well defined---we would need to define the state space of the full interacting quantum field theory to define them. Nevertheless, formally, one could argue that these formal Coulomb states should be orthogonal and that therefore Alice's particle is already decohered at the earliest time depicted in \cref{fig:originexp}. However this decoherence is a ``false decoherence'' in the sense of \cite{Unruh_2000}. If Alice recombines her particle slowly enough and if there are no external influences, she will be able to fully restore the coherence of her particle.

As Alice recombines her particle and moves its components along noninertial paths, formally the total state should continue to be of the form \cref{eq:Aliceinst2}. However, while the recombination process is occurring, there is no way to meaningfully separate $\ket{\psi_{1}}$ or $\ket{\psi_{2}}$ into a ``Coulomb part'' (which is not an independent degree of freedom and should cause only a false decoherence of Alice's particle) and a ``radiation part'' (which is a state of the free electromagnetic field that should be responsible for a true decoherence). Since we do not have a well-defined inner product between $\ket{\psi_{1}}$ and $\ket{\psi_{2}}$, we cannot, in general, meaningfully say how much true decoherence has occurred at any finite time during this process.

However, the situation improves considerably if we go to asymptotically late times. At asymptotically late times, the electromagnetic field naturally decomposes into a radiation field that propagates to null infinity and a Coulomb field that follows Alice's particle to timelike infinity. The asymptotic Coulomb field is completely determined by the asymptotic state of Alice's particle and does not represent an independent degree of freedom (see e.g. \cite{Satishchandran_2022}). Thus, at asymptotically late times, the state of the total system is of the form
\be 
\label{eq:Aliceinst3}
\frac{1}{\sqrt{2}}\big(\ket{\uparrow;A_{1}}_{i^+} \otimes \ket{\Psi_1}_{\scri^{+}}+\ket{\downarrow;A_{2}}_{i^+}  \otimes \ket{\Psi_2}_{\scri^{+}}\big).
\ee 
Here $\ket{\uparrow;A_{1}}_{i^+} $ and $\ket{\downarrow;A_{2}}_{i^+} $ represent the asymptotically late time states of the components of Alice's recombined particle and $\ket{\Psi_1}_{\scri^{+}}$ and $\ket{\Psi_2}_{\scri^{+}}$ represent the states of the radiation field at null infinity that would arise if, over all time, the states of Alice's particle were $\ket{\uparrow;A_{1}(t)}$ and $\ket{\downarrow;A_{2}(t)}$, respectively. 
Note that after recombination, the spatial wave packets describing the ``$1$'' and ``$2$'' states coincide, so, in particular, we have $\ket{A_{1}}_{i^+} = \ket{A_{2}}_{i^+}$, but we keep the $1$ and $2$ subscripts for notational clarity.

It is very important to recognize that---unlike \cref{eq:Aliceinst2}---\cref{eq:Aliceinst3} is not merely a formal expression. The states $\ket{\Psi_1}_{\scri^{+}}$ and $\ket{\Psi_2}_{\scri^{+}}$ are well-defined Fock space states of the ``out'' Hilbert space of the electromagnetic field and have a well-defined description in terms of photons.\footnote{In a general scattering process, there will be a nontrivial electromagnetic ``memory effect,'' resulting in infrared divergences in the description of the quantum state (see e.g. \cite{Satishchandran_2022,Kulish_1970,Bloch_1937}). In that case, the electromagnetic ``out'' state cannot be expressed as a state in the standard Fock space and cannot be given a proper description in terms of photons. However, such infrared divergences do not occur in cases where the charges are not relatively boosted at asymptotically early and late times as we consider here, so such infrared issues play no role in the analysis of this gedankenexperiment. Similar divergences which arise due to the gravitational memory effect also play no role in the (linearized) gravitational version of the gedankenexperiment.} The failure of $\ket{\Psi_1}_{\scri^{+}}$ and $\ket{\Psi_2}_{\scri^{+}}$ to coincide implies a decoherence of Alice's particle. The degree of decoherence of the asymptotic state of Alice's particle is given by 
\be 
\label{decAlicescri}
{\mathscr D}_{\text{Alice}} = 1 - \left\lvert\braket{\Psi_{1}|\Psi_{2}}_{\scri^{+}}\right\rvert
\ee 
where $\braket{\Psi_{1}|\Psi_{2}}_{\scri^{+}}$ denotes the inner product of the states $\ket{\Psi_{1}}_{\scri^{+}}$ and $\ket{\Psi_{2}}_{\scri^{+}}$ on $\scri^{+}$. This equation is a precise and general version of the decoherence estimate given in \cref{sec:review} based on the number of ``entangling photons'' that are emitted. If $\ket{\Psi_1}_{\scri^{+}}$ and $\ket{\Psi_2}_{\scri^{+}}$ differ by more than one photon, they should be nearly orthogonal, and the decoherence will be nearly complete.

\begin{figure}[h!]
    \centering
    \includegraphics[width=8.6cm]{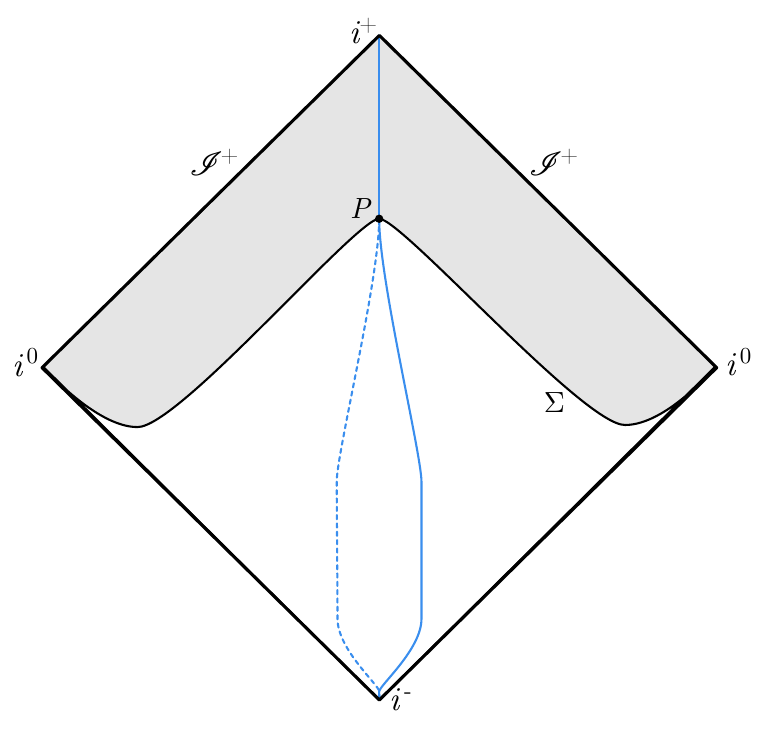}
    \caption{ Alice recombines her particle at event $P$ and subsequently keeps her recombined particle in inertial motion. $\Sigma$ is an arbitrary Cauchy surface passing through $P$.
    }
    \label{fig:gedanken1}
\end{figure}

We are interested in the case depicted in \cref{fig:gedanken1} where Alice recombines her particle as in the gedankenexperiment---but without the presence of Bob---and after recombination, she keeps her combined particle in inertial motion at all future times.  Then, to the causal future of the recombination event $P$, the electromagnetic field will correspond to the Coulomb field of the recombined particle. Let $\Sigma$ be an arbitrary Cauchy surface passing through $P$. Extend the Coulomb field of the recombined particle to the entire region to the future, $I^+(\Sigma)$, of $\Sigma$ (i.e., not just the causal future of $P$). Subtract this Coulomb field from the electromagnetic field in this region. The electromagnetic field associated with $\ket{\uparrow;A_{1}}$ with the final Coulomb field subtracted will thus correspond to a well-defined state $\ket{\Psi_{1}}_\Sigma$ of the source-free electromagnetic field on $\Sigma$. Similarly, the electromagnetic field associated with $\ket{\downarrow;A_{2}}$ with the final Coulomb field subtracted will correspond to a well-defined state $\ket{\Psi_{2}}_\Sigma$ on $\Sigma$. At ``time'' $\Sigma$, the joint state of Alice's particle and the electromagnetic field is described by
\be 
\label{eq:Aliceinst10}
\frac{1}{\sqrt{2}}\big(\ket{\uparrow;A_{1}}_\Sigma \otimes \ket{\Psi_1}_{\Sigma}+\ket{\downarrow;A_{2}}_{\Sigma}  \otimes \ket{\Psi_2}_{\Sigma}\big).
\ee
In contrast to \cref{eq:Aliceinst2}, this is a completely meaningful expression; $\ket{\Psi_{1}}_\Sigma$ and $\ket{\Psi_{2}}_\Sigma$ are well-defined states of the source-free electromagnetic field. Under time evolution, $\ket{\Psi_{1}}_\Sigma$ and $\ket{\Psi_{2}}_\Sigma$ evolve to $\ket{\Psi_1}_{\scri^{+}}$ and $\ket{\Psi_2}_{\scri^{+}}$, respectively. Since time evolution is unitary, we may express the decoherence \cref{decAlicescri} of Alice's particle as 
\begin{equation}
\label{eq:decAliceSigma0}
{\mathscr D}_{\text{Alice}} = 1 - |\braket{\Psi_{1}|\Psi_{2}}_{\Sigma}|.
\end{equation}
This is our desired expression for the decoherence due to Alice. It is clear that if there are no time constraints on Alice's recombination, then by doing the recombination adiabatically---so that negligible radiation is emitted to infinity---she can make the decoherence arbitrarily small.

\subsection{Decoherence Due to Bob}
\label{subsec:decBob}

We now consider the decoherence that would occur if Bob makes a measurement that obtains some which-path information about Alice's particle. We assume that Alice recombines her particle adiabatically in the distant future---after Bob has completed his measurements---in such a way that had Bob not been present, no decoherence would have occurred. Thus, any decoherence in this situation can be attributed to Bob. This situation corresponds to experimental proposals such as \cite{Bose_2017}.

Since Bob is now part of the system, heuristically, the state of the total system after Alice has put her particle through the initial Stern-Gerlach apparatus but before Bob has begun his measurements is now
\be 
\label{eq:Aliceinst4}
\frac{1}{\sqrt{2}}\big(\ket{\uparrow;A_{1}} \otimes \ket{\psi_1}+\ket{\downarrow;A_{2}} \otimes \ket{\psi_2}\big) \otimes \ket{B_0}
\ee 
where $\ket{B_0}$ is the initial state of Bob's apparatus and again $\ket{\psi_{1}}$ and $\ket{\psi_{2}}$ are the formal Coulomb states of Alice's particle. We wish to consider a situation wherein Bob turns on his apparatus for a time $T_B$ and makes a measurement of the Coulomb field of Alice's particle in order to try to obtain which-path information. We assume that Bob carries out his measurement in such a way that he emits negligible radiation to infinity. For example if Bob measures the motion of a charged particle released from a trap as described in the previous section, the sensitivity of his experiment will depend on $q_B/m_B$ but the emitted radiation will vary as $q^2_B$, so by taking $q_B$ and $m_B$ sufficiently small, he should be able to carry out his measurements with negligible emitted radiation.\footnote{The assumption that Bob emits negligible radiation is being made so as to make our discussion simpler and cleaner, but it is not essential for the analysis.} We allow Bob to make any field measurement whatsoever, i.e., we do not restrict him to measuring the trajectory of a particle released from a trap. For the analysis of this subsection, we do not place any limits on $T_B$, i.e., we do not require $T_B < D$.

Since no radiation is emitted by Bob or Alice, at asymptotically late times, the state of the electromagnetic field at null infinity will be $\ket{0}_{\scri^+}$ for either state of Alice's superposition. Thus, the final state of the electromagnetic field plays no role in entanglement and we need only be concerned with the Alice-Bob system. The final state of the Alice-Bob system will be of the form
\be 
\label{eq:Aliceinst5}
\frac{1}{\sqrt{2}}\big(\ket{\uparrow;A_{1}}_{i^+} \otimes \ket{B_1}_{i^{+}}+\ket{\downarrow;A_{2}}_{i^+}  \otimes \ket{B_2}_{i^{+}}\big)
\ee
where $\ket{B_1}_{i^{+}}$ and $\ket{B_2}_{i^{+}}$ are the final states of Bob's apparatus for Alice's states $\ket{\uparrow;A_{1}}$ and $\ket{\downarrow;A_{2}}$, respectively. The failure of $\ket{B_1}_{i^{+}}$ and $\ket{B_2}_{i^{+}}$ to coincide corresponds to Bob having acquired which-path information about Alice's particle. The corresponding decoherence of Alice's particle is
\be 
\label{decBobi}
{\mathscr D}_{\text{Bob}} = 1 - |\braket{B_{1}|B_{2}}_{i^{+}}|.
\ee 
However, since Bob stops interacting at time $T_B$, we can equivalently calculate the inner product at time $T_B$
\be 
\label{decBobTB}
{\mathscr D}_{\text{Bob}} = 1 - |\braket{B_{1}|B_{2}}_{T_B}|.
\ee 
This gives the decoherence associated with Bob's measurement. In the circumstance considered here where Alice emits no radiation, it is clear that this decoherence can be viewed as being caused by Bob. It also is clear that in this circumstance, the decoherence should be viewed as being mediated by the Coulomb field of Alice's particle.

\Cref{decBobTB} is a precise and general version of the decoherence estimate given in \cref{sec:review} based upon Bob's ability to get which-path information. The amount of which-path information Bob can obtain is determined by the extent to which Bob can design a measurement so that $\ket{B_1}_{T_B}$ is nearly orthogonal to $\ket{B_2}_{T_B}$. The degree to which $\ket{B_1}_{T_B}$ is orthogonal to $\ket{B_2}_{T_B}$ determines how much decoherence of Alice's particle must occur.

\section{Reanalysis of the gedankenexperiment}
\label{sec:gedankres}

\begin{figure}[h]
    \includegraphics[width=\linewidth]{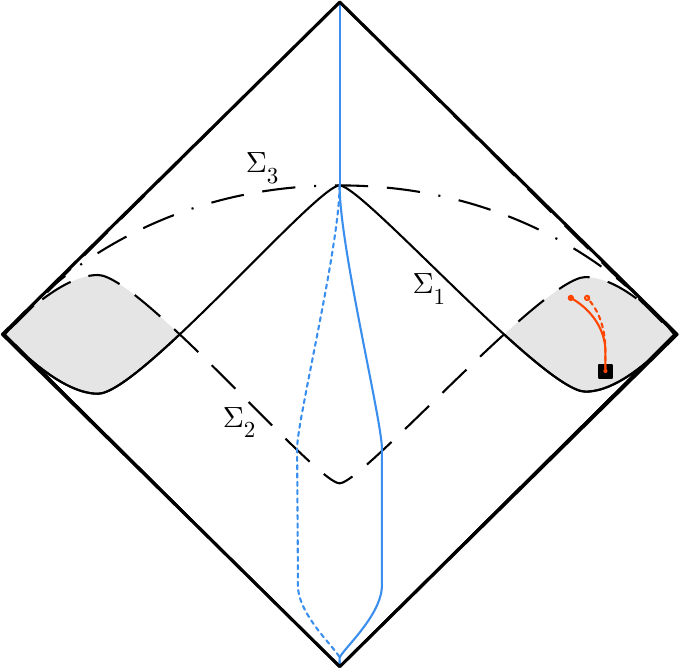}
    \caption{
    A spacetime diagram of the gedankenexperiment of \cref{fig:originexp} showing the three Cauchy surfaces, $\Sigma_1$, $\Sigma_2$, and $\Sigma_3$. The Cauchy surface $\Sigma_1$ passes through Alice's region after recombination but is such that the region in which Bob performs his measurements (shaded in gray) lies to the future of $\Sigma_1$. (We have depicted Bob as releasing a particle from a trap, but Bob is allowed to perform any measurement whatsoever in the gray region.) The Cauchy surface $\Sigma_2$ is such that it passes through Alice's region before she starts the recombination process but is such that Bob's measurement lies to the past of $\Sigma_2$. The Cauchy surface $\Sigma_3$ passes through Alice's region after recombination and is such that Bob's measurement lies to the past of $\Sigma_3$.}
    \label{fig:gedanken}
\end{figure}

We now are in a position to reanalyze the gedankenexperiment of \cref{sec:review}. We will again explicitly consider the electromagnetic version of the gedankenexperiment, but the exactly same discussion applies to the gravitational case with the appropriate word substitutions. The spacetime diagram of the gedankenexperiment is redrawn in \cref{fig:gedanken} in order to show three Cauchy surfaces, $\Sigma_1$, $\Sigma_2$, and $\Sigma_3$, that will play an important role in our reanalysis.

We reanalyze the decoherence of Alice's particle using the results of the previous section as follows. First, consider the portion of the spacetime of \cref{fig:gedanken} that lies to the past of Cauchy surface $\Sigma_1$. At the time represented by $\Sigma_1$, Alice has completed her recombination but Bob has not yet begun performing his measurements. The portion of the spacetime lying to the past of $\Sigma_1$ is identical to the portion of the spacetime of \cref{fig:gedanken1} lying to the past of a corresponding Cauchy surface $\Sigma$. Thus, we may apply the results of \cref{subsec:decAlice} to conclude that the decoherence of Alice's particle is given by
\begin{equation}
\label{eq:decAliceSigma1}
{\mathscr D}_{\text{Alice}} = 1 - \left\lvert\braket{\Psi_{1}|\Psi_{2}}_{\Sigma_1}\right\rvert,
\end{equation}
where $\ket{\Psi_{1}}_{\Sigma_1}$ and $\ket{\Psi_{2}}_{\Sigma_1}$ are the radiation states on $\Sigma_1$ obtained by subtracting the common Coulomb field from the states of the electromagnetic field corresponding to Alice's particle being in states $\ket{\uparrow;A_{1}}$ and $\ket{\downarrow;A_{2}}$, respectively. Since Alice's recombination is complete at time $\Sigma_1$, \cref{eq:decAliceSigma1} should yield the exact expression for the decoherence of Alice's particle.

However, we also can analyze the decoherence of Alice's particle by considering the portion of the spacetime that lies to the past of the Cauchy surface $\Sigma_2$. At time $\Sigma_2$, Alice has not yet started her recombination, but Bob has completed his measurements. Thus, the situation here is identical to the setup considered in \cref{subsec:decBob}. Hence, we may apply the results of \cref{subsec:decBob} to conclude that a decoherence of Alice's particle given by
\be 
\label{decBob2}
{\mathscr D}_{\text{Bob}} = 1 - |\braket{B_{1}|B_{2}}|
\ee 
must occur as a result of Bob's measurements, where $\ket{B_1}$ and $\ket{B_2}$ represent the states of Bob's apparatus after completion of his measurement. It is possible that more decoherence of Alice's particle could occur as Alice performs her recombination. However, since Bob has completed his measurement and stops interacting after time $\Sigma_2$, it is impossible for the decoherence of Alice's particle to be less than this.

It follows that there would be a paradox if it were possible for Bob to do a measurement in such a way that
\be
|\braket{B_{1}|B_{2}}| < \left\lvert\braket{\Psi_{1}|\Psi_{2}}_{\Sigma_1}\right\rvert,
\label{decparineq}
\ee
i.e., such that the decoherence associated with Bob's measurement is greater than the decoherence due to Alice. If \cref{decparineq} held, then Bob's measurement either would result in a violation of causality [if it induced an additional decoherence of Alice's particle beyond that given by \cref{eq:decAliceSigma1}], or it would result in a violation of complementarity (if it did not induce such an additional decoherence). \cref{decparineq} is a precise statement of the potential paradox posed by the gedankenexperiment of \cref{sec:review}.

However, it is now easy to see that no such paradox can ever arise. At time $\Sigma_1$, the state of the joint Alice-field-Bob system is described by
\be
\label{eq:Aliceinst7}
\frac{1}{\sqrt{2}}\big(\ket{\uparrow;A_{1}} \otimes \ket{\Psi_1}_{\Sigma_1}+\ket{\downarrow;A_{2}} \otimes \ket{\Psi_2}_{\Sigma_1} \big) \otimes \ket{B_0}
\ee 
where $\ket{\Psi_{1}}_{\Sigma_1}$ and $\ket{\Psi_{2}}_{\Sigma_1}$ are the radiation states on $\Sigma_1$ (with the common Coulomb field subtracted), and $\ket{B_0}$ is the initial state of Bob's detector. We now consider the evolution of this state to the Cauchy surface $\Sigma_3$. There is no evolution of Alice's state, since $\Sigma_3$ is the same time as $\Sigma_1$ as far as Alice's state is concerned. However, the radiation interacts with Bob's measuring apparatus. In the case where Alice's state is $\ket{\uparrow;A_{1}}$, Bob's state evolves to $\ket{B_1}$, whereas if Alice's state is $\ket{\downarrow;A_{2}}$, Bob's state evolves to $\ket{B_2}$. It follows that the state \cref{eq:Aliceinst7} on $\Sigma_1$ must evolve to the state on $\Sigma_3$ described by
\be
\label{eq:Aliceinst9}
\frac{1}{\sqrt{2}}\big(\ket{\uparrow;A_{1}} \otimes \ket{\Psi'_1}_{\Sigma_3} \otimes \ket{B_1}+\ket{\downarrow;A_{2}} \otimes \ket{\Psi'_2}_{\Sigma_3} \otimes \ket{B_2} \big).
\ee 
Here  $\ket{\Psi'_1}_{\Sigma_3}$ and $ \ket{\Psi'_2}_{\Sigma_3}$ are the radiation states that arise from $\ket{\Psi_{1}}_{\Sigma_1}$ and $\ket{\Psi_{2}}_{\Sigma_1}$, respectively, after interaction with Bob. The states $\ket{\Psi'_1}_{\Sigma_3}$ and $ \ket{\Psi'_2}_{\Sigma_3}$ depend on the interaction with Bob, so they cannot be calculated without knowing exactly what Bob is measuring. However, no matter what Bob does, the joint evolution from $\Sigma_1$ to $\Sigma_3$ must be unitary. It follows that the norms of states are preserved and that
\begin{eqnarray}
\braket{\Psi'_{1}|\Psi'_{2}}_{\Sigma_3} \braket{B_1|B_2} &=& \braket{\Psi_{1}|\Psi_{2}}_{\Sigma_1} \braket{B_0|B_0} \nonumber \\
&=& \braket{\Psi_{1}|\Psi_{2}}_{\Sigma_1}. 
\end{eqnarray}
It then follows immediately that 
\be
|\braket{B_{1}|B_{2}}|  \geq \left\lvert\braket{\Psi_{1}|\Psi_{2}}_{\Sigma_1}\right\rvert
\label{decparineq2}
\ee
so the inequality \cref{decparineq} can never be satisfied. This is precisely what we wished to show.

Although the above argument completes our proof that no contradiction with causality or complementarity can ever arise in this gedankenexperiment---no matter what Bob chooses to measure---it remains to give a more intuitive explanation of our new resolution of the gedankenexperiment and connect it with the discussion of \cref{sec:review}. 

The main new ingredient that we have added to the analysis is that we may view Bob as measuring aspects of the radiation emitted by Alice's particle. It may seem strange to talk about ``emitted radiation'' that is present in a region that is spacelike separated from the region where the emission is taking place. Indeed, this may, by itself, appear to be a violation of causality! However, this kind of phenomenon is a basic feature of quantum field theory, with no violation of causality involved. The mode function of a particle in quantum field theory is a positive frequency solution and cannot be sharply localized. If a photon is emitted by a source in some localized region $\mathcal O$, there always will be some amplitude for the photon to be present in a region spacelike separated from $\mathcal O$. Indeed, as discussed in detail in \cite{Unruh_1984}, there are cases where the emitted photon is {\em mostly} localized in a spacelike separated region. This does not lead to a violation of causality because an observer in the spacelike separated region will not be able to tell whether she is observing a photon or a vacuum fluctuation---she can tell the difference between these possibilities only when she enters the causal future of $\mathcal O$. In the present case, the electromagnetic field in Bob's region can be viewed either as corresponding to the superposition of the Coulomb fields of Alice's particle with no radiation---as would be natural to do if we view Bob's region as lying to the past of time $\Sigma_2$---or as the single Coulomb field of Alice's combined particle together with free radiation---as would be natural to do if we view Bob's region as lying to the future of time $\Sigma_1$. These viewpoints are indistinguishable in Bob's region.

The radiation viewpoint allows us to understand why Bob cannot produce any additional decoherence beyond what Alice produces during her recombination. Bob can obtain which-path information only by measuring (i.e., scattering and/or absorbing) the entangling photons that ``previously'' were emitted by Alice. Therefore, the state of his apparatus cannot become more correlated with Alice's particle than the radiation emitted by Alice, as we have proven above in \cref{decparineq2}.

Note that, as we have just argued, in the gedankenexperiment, Bob is merely an ``innocent bystander'' with regard to the decoherence of Alice's particle, since he is merely measuring the entangling radiation emitted by the particle that was the true cause of the decoherence. However, suppose that Alice does not follow the protocol assigned to her in the gedankenexperiment and instead recombines her particle very slowly at a later time, so as not to produce any radiation. Then, despite her attempts to keep perfect coherence, she will find that her particle has decohered by the amount \cref{decBob2}. In this case, Bob's measurement is the true cause of her particle's decoherence \cite{Belenchia:2019gcc}. Interestingly, when Bob performs his measurements, he has no way of knowing whether he will turn out to be an ``innocent bystander'' or the cause of decoherence of Alice's particle.

Finally, we note that the analysis of the gedankenexperiment summarized in \cref{sec:review} was based upon the limitations on Alice's ability to maintain coherence due to radiation and the limitations on Bob's ability to get which-path information due to vacuum fluctuations. The reanalysis of the gedankenexperiment given above gave a more precise version of Alice's limitations on maintaining coherence due to radiation. However, we did not mention ``vacuum fluctuations'' in the discussion of the decoherence associated with Bob's measurements, so it might appear that the reanalysis differs in this respect. However, this is not the case: The radiation fields $\ket{\Psi_{1}}_{\Sigma_1}$ and $\ket{\Psi_{2}}_{\Sigma_1}$ have different expected values of the electromagnetic field. Their failure to be orthogonal can be viewed as a manifestation of the same type of fluctuations in these states as occurs in the vacuum state; if these states did not have such fluctuations, they would be fully distinguishable and hence orthogonal. But, as is evident from \ref{decparineq2}, it is the failure of $\ket{\Psi_{1}}_{\Sigma_1}$ and $\ket{\Psi_{2}}_{\Sigma_1}$ to be orthogonal that limits Bob's ability to make $\ket{B_1}$ and $\ket{B_2}$ orthogonal. Thus, there is a direct connection between vacuum fluctuations and the limitations on Bob's ability to obtain which-path information.

\section{Summary and Conclusions}
\label{sec:summconc}

In this paper, we have reanalyzed the gedankenexperiment discussed in \cite{Wald2018}. Our reanalysis validates the arguments that had been made in \cite{Wald2018} using only back-of-the-envelope estimates, and it shows in a much more precise way---and under completely general assumptions about the measurements that Bob makes---that no violations of causality or complementarity can occur.

Perhaps the most interesting aspect of our reanalysis is the equivalence of two viewpoints on how the state of Bob's measuring apparatus becomes correlated with the state of Alice's particle. In the gravitational version of the gedankenexperiment, one can say either that (i) Alice's particle became entangled with on-shell gravitons emitted during the recombination process and Bob's apparatus then interacted with these gravitons---thereby transferring some of the entanglement present in these gravitons to his apparatus---or that (ii) the Newtonian gravitational field of Alice's particle mediated an entanglement of Bob's apparatus with Alice's particle. If Alice follows her protocol but Bob fails to make any measurement, then it is essential to take viewpoint (i) to understand why Bob's inaction has no effect whatsoever on the decoherence of Alice's particle. Conversely, if Bob follows his protocol but Alice recombines her particle adiabatically at a later time, one must take viewpoint (ii) to understand how Bob's measuring apparatus becomes correlated with the Alice's particle \cite{Belenchia:2019gcc}. But if Alice and Bob each follow the protocols of the gedankenexperiment, then both (i) and (ii) provide a valid description of the process that occurs. 

Indeed, it is essential that both (i) and (ii)---or, alternatively, neither (i) nor (ii)---be valid descriptions of the process. To see this, suppose that (i) fails, i.e., Alice's particle does not emit entangling gravitons, but suppose that (ii) holds, i.e., Bob's apparatus is able to entangle with Alice's particle via its Newtonian gravitational field. Then Alice's particle would not decohere in the absence of Bob. It follows that if it decohered in the presence of Bob we would have a violation of causality, whereas if it did not decohere in the presence of Bob we would have a violation of complementarity. Thus, it is not consistent for (i) to fail but (ii) to hold. Conversely, suppose (i) holds, i.e., Alice's particle emits quantized entangling gravitational radiation, but suppose that (ii) fails, i.e., Bob's apparatus is unable to entangle with Alice's particle via its Newtonian gravitational field. Then, since, as we have seen, under the protocol of the gedankenexperiment, the difference of the Newtonian fields of Alice's particle can be equivalently viewed as quantized radiation emitted by Alice's particle, this would imply that Bob is unable to interact with quantized gravitational radiation in any way that results in entanglement. This would not make sense in any theory where quantized gravitational radiation can be produced.\footnote{It has been argued that it may be impossible, in principle, to measure the energy of a single graviton \cite{DYSON_2013}. Bob is not required here to resolve an individual graviton but merely to become entangled, at least to some degree, with gravitons.}

These considerations show that there is a direct relationship between Newtonian entanglement and the existence of gravitons. Our argument for such a relationship is strictly valid only within the protocol of the gedankenexperiment, where the measurement of the Newtonian field/gravitons is carried out within a time span no longer than the light travel time to the source. Nevertheless, these considerations yield strong support for the view that any observation of entanglement mediated by a Newtonian field provides evidence for the existence of the graviton.

\begin{acknowledgments}
D.L.D. acknowledges support as a Fannie and John Hertz Foundation Fellow holding the Barbara Ann Canavan Fellowship and as an Eckhardt Graduate Scholar in the Physical Sciences Division at the University of Chicago. This research was supported in part by NSF Grant No. 21-05878 to the University of Chicago.
\end{acknowledgments}
\bibliography{arXiv_reupload}

\begin{thebibliography}{49}%
\makeatletter
\providecommand \@ifxundefined [1]{%
 \@ifx{#1\undefined}
}%
\providecommand \@ifnum [1]{%
 \ifnum #1\expandafter \@firstoftwo
 \else \expandafter \@secondoftwo
 \fi
}%
\providecommand \@ifx [1]{%
 \ifx #1\expandafter \@firstoftwo
 \else \expandafter \@secondoftwo
 \fi
}%
\providecommand \natexlab [1]{#1}%
\providecommand \enquote  [1]{``#1''}%
\providecommand \bibnamefont  [1]{#1}%
\providecommand \bibfnamefont [1]{#1}%
\providecommand \citenamefont [1]{#1}%
\providecommand \href@noop [0]{\@secondoftwo}%
\providecommand \href [0]{\begingroup \@sanitize@url \@href}%
\providecommand \@href[1]{\@@startlink{#1}\@@href}%
\providecommand \@@href[1]{\endgroup#1\@@endlink}%
\providecommand \@sanitize@url [0]{\catcode `\\12\catcode `\$12\catcode
  `\&12\catcode `\#12\catcode `\^12\catcode `\_12\catcode `\%12\relax}%
\providecommand \@@startlink[1]{}%
\providecommand \@@endlink[0]{}%
\providecommand \url  [0]{\begingroup\@sanitize@url \@url }%
\providecommand \@url [1]{\endgroup\@href {#1}{\urlprefix }}%
\providecommand \urlprefix  [0]{URL }%
\providecommand \Eprint [0]{\href }%
\providecommand \doibase [0]{https://doi.org/}%
\providecommand \selectlanguage [0]{\@gobble}%
\providecommand \bibinfo  [0]{\@secondoftwo}%
\providecommand \bibfield  [0]{\@secondoftwo}%
\providecommand \translation [1]{[#1]}%
\providecommand \BibitemOpen [0]{}%
\providecommand \bibitemStop [0]{}%
\providecommand \bibitemNoStop [0]{.\EOS\space}%
\providecommand \EOS [0]{\spacefactor3000\relax}%
\providecommand \BibitemShut  [1]{\csname bibitem#1\endcsname}%
\let\auto@bib@innerbib\@empty
\bibitem [{\citenamefont {Belenchia}\ \emph {et~al.}(2018)\citenamefont
  {Belenchia}, \citenamefont {Wald}, \citenamefont {Giacomini}, \citenamefont
  {Castro-Ruiz}, \citenamefont {Brukner},\ and\ \citenamefont
  {Aspelmeyer}}]{Wald2018}%
  \BibitemOpen
  \bibfield  {author} {\bibinfo {author} {\bibfnamefont {A.}~\bibnamefont
  {Belenchia}}, \bibinfo {author} {\bibfnamefont {R.~M.}\ \bibnamefont {Wald}},
  \bibinfo {author} {\bibfnamefont {F.}~\bibnamefont {Giacomini}}, \bibinfo
  {author} {\bibfnamefont {E.}~\bibnamefont {Castro-Ruiz}}, \bibinfo {author}
  {\bibfnamefont {v.}~\bibnamefont {Brukner}},\ and\ \bibinfo {author}
  {\bibfnamefont {M.}~\bibnamefont {Aspelmeyer}},\ }\bibfield  {title}
  {\bibinfo {title} {Quantum superposition of massive objects and the
  quantization of gravity},\ }\href
  {https://doi.org/10.1103/PhysRevD.98.126009} {\bibfield  {journal} {\bibinfo
  {journal} {Phys. Rev. D}\ }\textbf {\bibinfo {volume} {98}},\ \bibinfo
  {pages} {126009} (\bibinfo {year} {2018})},\ \Eprint
  {https://arxiv.org/abs/1807.07015} {arXiv:1807.07015 [quant-ph]} \BibitemShut
  {NoStop}%
\bibitem [{\citenamefont {Belenchia}\ \emph {et~al.}(2019)\citenamefont
  {Belenchia}, \citenamefont {Wald}, \citenamefont {Giacomini}, \citenamefont
  {Castro-Ruiz}, \citenamefont {Brukner},\ and\ \citenamefont
  {Aspelmeyer}}]{Belenchia:2019gcc}%
  \BibitemOpen
  \bibfield  {author} {\bibinfo {author} {\bibfnamefont {A.}~\bibnamefont
  {Belenchia}}, \bibinfo {author} {\bibfnamefont {R.~M.}\ \bibnamefont {Wald}},
  \bibinfo {author} {\bibfnamefont {F.}~\bibnamefont {Giacomini}}, \bibinfo
  {author} {\bibfnamefont {E.}~\bibnamefont {Castro-Ruiz}}, \bibinfo {author}
  {\bibfnamefont {v.}~\bibnamefont {Brukner}},\ and\ \bibinfo {author}
  {\bibfnamefont {M.}~\bibnamefont {Aspelmeyer}},\ }\bibfield  {title}
  {\bibinfo {title} {Information content of the gravitational field of a
  quantum superposition},\ }\href {https://doi.org/10.1142/S0218271819430016}
  {\bibfield  {journal} {\bibinfo  {journal} {Int. J. Mod. Phys. D}\ }\textbf
  {\bibinfo {volume} {28}},\ \bibinfo {pages} {1943001} (\bibinfo {year}
  {2019})},\ \Eprint {https://arxiv.org/abs/1905.04496} {arXiv:1905.04496
  [quant-ph]} \BibitemShut {NoStop}%
\bibitem [{\citenamefont {Bronstein}(2012)}]{Bronstein}%
  \BibitemOpen
  \bibfield  {author} {\bibinfo {author} {\bibfnamefont {M.}~\bibnamefont
  {Bronstein}},\ }\bibfield  {title} {\bibinfo {title} {Republication of:
  Quantum theory of weak gravitational fields},\ }\href
  {https://doi.org/10.1007/s10714-011-1285-4} {\bibfield  {journal} {\bibinfo
  {journal} {General Relativity and Gravitation}\ }\textbf {\bibinfo {volume}
  {44}},\ \bibinfo {pages} {267} (\bibinfo {year} {2012})}\BibitemShut
  {NoStop}%
\bibitem [{\citenamefont {Page}\ and\ \citenamefont
  {Geilker}(1981)}]{Page_1981}%
  \BibitemOpen
  \bibfield  {author} {\bibinfo {author} {\bibfnamefont {D.~N.}\ \bibnamefont
  {Page}}\ and\ \bibinfo {author} {\bibfnamefont {C.~D.}\ \bibnamefont
  {Geilker}},\ }\bibfield  {title} {\bibinfo {title} {Indirect evidence for
  quantum gravity},\ }\href {https://doi.org/10.1103/PhysRevLett.47.979}
  {\bibfield  {journal} {\bibinfo  {journal} {Phys. Rev. Lett.}\ }\textbf
  {\bibinfo {volume} {47}},\ \bibinfo {pages} {979} (\bibinfo {year}
  {1981})}\BibitemShut {NoStop}%
\bibitem [{\citenamefont {Eppley}\ and\ \citenamefont
  {Hannah}(1977)}]{Eppley_1977}%
  \BibitemOpen
  \bibfield  {author} {\bibinfo {author} {\bibfnamefont {K.}~\bibnamefont
  {Eppley}}\ and\ \bibinfo {author} {\bibfnamefont {E.}~\bibnamefont
  {Hannah}},\ }\bibfield  {title} {\bibinfo {title} {The necessity of
  quantizing the gravitational field},\ }\href
  {https://doi.org/10.1007/BF00715241} {\bibfield  {journal} {\bibinfo
  {journal} {Foundations of Physics}\ }\textbf {\bibinfo {volume} {7}},\
  \bibinfo {pages} {51} (\bibinfo {year} {1977})}\BibitemShut {NoStop}%
\bibitem [{\citenamefont {Mattingly}(2006)}]{Mattingly_2006}%
  \BibitemOpen
  \bibfield  {author} {\bibinfo {author} {\bibfnamefont {J.}~\bibnamefont
  {Mattingly}},\ }\bibfield  {title} {\bibinfo {title} {{Why Eppley and
  Hannah's thought experiment fails}},\ }\href
  {https://doi.org/10.1103/PhysRevD.73.064025} {\bibfield  {journal} {\bibinfo
  {journal} {Phys. Rev. D}\ }\textbf {\bibinfo {volume} {73}},\ \bibinfo
  {pages} {064025} (\bibinfo {year} {2006})},\ \Eprint
  {https://arxiv.org/abs/0601127} {arXiv:0601127} \BibitemShut {NoStop}%
\bibitem [{\citenamefont {Carlip}(2008)}]{Carlip_2008}%
  \BibitemOpen
  \bibfield  {author} {\bibinfo {author} {\bibfnamefont {S.}~\bibnamefont
  {Carlip}},\ }\bibfield  {title} {\bibinfo {title} {Is quantum gravity
  necessary?},\ }\href {https://doi.org/10.1088/0264-9381/25/15/154010}
  {\bibfield  {journal} {\bibinfo  {journal} {Class. Quant. Grav.}\ }\textbf
  {\bibinfo {volume} {25}},\ \bibinfo {pages} {154010} (\bibinfo {year}
  {2008})},\ \Eprint {https://arxiv.org/abs/0803.3456} {arXiv:0803.3456
  [gr-qc]} \BibitemShut {NoStop}%
\bibitem [{\citenamefont {Giampaolo}\ and\ \citenamefont
  {Macr\`\i{}}(2019)}]{Giampaolo_2019}%
  \BibitemOpen
  \bibfield  {author} {\bibinfo {author} {\bibfnamefont {S.~M.}\ \bibnamefont
  {Giampaolo}}\ and\ \bibinfo {author} {\bibfnamefont {T.}~\bibnamefont
  {Macr\`\i{}}},\ }\bibfield  {title} {\bibinfo {title} {Entanglement,
  holonomic constraints, and the quantization of fundamental interactions},\
  }\href {https://doi.org/10.1038/s41598-019-47844-8} {\bibfield  {journal}
  {\bibinfo  {journal} {Sci. Rep.}\ }\textbf {\bibinfo {volume} {9}},\ \bibinfo
  {pages} {11362} (\bibinfo {year} {2019})},\ \Eprint
  {https://arxiv.org/abs/1806.08383} {arXiv:1806.08383 [quant-ph]} \BibitemShut
  {NoStop}%
\bibitem [{\citenamefont {Hossenfelder}(2010)}]{Hossenfelder:2010}%
  \BibitemOpen
  \bibfield  {author} {\bibinfo {author} {\bibfnamefont {S.}~\bibnamefont
  {Hossenfelder}},\ }\bibinfo {title} {{Experimental Search for Quantum
  Gravity}},\ in\ \href@noop {} {\emph {\bibinfo {booktitle} {{Classical and
  quantum gravity}: {Theory, Analysis and Applications}}}}\ (\bibinfo
  {publisher} {Nova Science Publishers},\ \bibinfo {address} {Hauppauge, NY},\
  \bibinfo {year} {2010})\ \Eprint {https://arxiv.org/abs/1010.3420}
  {arXiv:1010.3420 [gr-qc]} \BibitemShut {NoStop}%
\bibitem [{\citenamefont {Penrose}(2014)}]{Penrose_2014}%
  \BibitemOpen
  \bibfield  {author} {\bibinfo {author} {\bibfnamefont {R.}~\bibnamefont
  {Penrose}},\ }\bibfield  {title} {\bibinfo {title} {On the gravitization of
  quantum mechanics 1: Quantum state reduction},\ }\href
  {https://doi.org/10.1007/s10701-013-9770-0} {\bibfield  {journal} {\bibinfo
  {journal} {Foundations of Physics}\ }\textbf {\bibinfo {volume} {44}},\
  \bibinfo {pages} {557} (\bibinfo {year} {2014})}\BibitemShut {NoStop}%
\bibitem [{\citenamefont {Diósi}(1987)}]{DIOSI_1987}%
  \BibitemOpen
  \bibfield  {author} {\bibinfo {author} {\bibfnamefont {L.}~\bibnamefont
  {Diósi}},\ }\bibfield  {title} {\bibinfo {title} {A universal master
  equation for the gravitational violation of quantum mechanics},\ }\href
  {https://doi.org/https://doi.org/10.1016/0375-9601(87)90681-5} {\bibfield
  {journal} {\bibinfo  {journal} {Physics Letters A}\ }\textbf {\bibinfo
  {volume} {120}},\ \bibinfo {pages} {377} (\bibinfo {year}
  {1987})}\BibitemShut {NoStop}%
\bibitem [{\citenamefont {Dyson}(2013)}]{DYSON_2013}%
  \BibitemOpen
  \bibfield  {author} {\bibinfo {author} {\bibfnamefont {F.}~\bibnamefont
  {Dyson}},\ }\bibfield  {title} {\bibinfo {title} {Is a graviton
  detectable?},\ }\href {https://doi.org/10.1142/S0217751X1330041X} {\bibfield
  {journal} {\bibinfo  {journal} {International Journal of Modern Physics A}\
  }\textbf {\bibinfo {volume} {28}},\ \bibinfo {pages} {1330041} (\bibinfo
  {year} {2013})}\BibitemShut {NoStop}%
\bibitem [{\citenamefont {DeWitt-Morette}\ and\ \citenamefont
  {Rickles}(2011)}]{Dewitt_2011}%
  \BibitemOpen
  \bibfield  {author} {\bibinfo {author} {\bibfnamefont {C.}~\bibnamefont
  {DeWitt-Morette}}\ and\ \bibinfo {author} {\bibfnamefont {D.}~\bibnamefont
  {Rickles}},\ }\bibfield  {title} {\bibinfo {title} {The role of gravitation
  in physics: report from the 1957 chapel hill conference}\ }(\bibinfo
  {publisher} {Max Planck Institute for the History of Science},\ \bibinfo
  {address} {Berlin},\ \bibinfo {year} {2011})\BibitemShut {NoStop}%
\bibitem [{\citenamefont {Zeh}(2011)}]{Zeh_2008}%
  \BibitemOpen
  \bibfield  {author} {\bibinfo {author} {\bibfnamefont {H.~D.}\ \bibnamefont
  {Zeh}},\ }\bibfield  {title} {\bibinfo {title} {Feynman's interpretation of
  quantum theory},\ }\href {https://doi.org/10.1140/epjh/e2011-10035-2}
  {\bibfield  {journal} {\bibinfo  {journal} {Eur. Phys. J. H}\ }\textbf
  {\bibinfo {volume} {36}},\ \bibinfo {pages} {63} (\bibinfo {year} {2011})},\
  \Eprint {https://arxiv.org/abs/0804.3348} {arXiv:0804.3348 [quant-ph]}
  \BibitemShut {NoStop}%
\bibitem [{\citenamefont {Gerlich}\ \emph {et~al.}(2011)\citenamefont
  {Gerlich}, \citenamefont {Eibenberger}, \citenamefont {Tomandl},
  \citenamefont {Nimmrichter}, \citenamefont {Hornberger}, \citenamefont
  {Fagan}, \citenamefont {T{\"u}xen}, \citenamefont {Mayor},\ and\
  \citenamefont {Arndt}}]{Gerlich_2011}%
  \BibitemOpen
  \bibfield  {author} {\bibinfo {author} {\bibfnamefont {S.}~\bibnamefont
  {Gerlich}}, \bibinfo {author} {\bibfnamefont {S.}~\bibnamefont
  {Eibenberger}}, \bibinfo {author} {\bibfnamefont {M.}~\bibnamefont
  {Tomandl}}, \bibinfo {author} {\bibfnamefont {S.}~\bibnamefont
  {Nimmrichter}}, \bibinfo {author} {\bibfnamefont {K.}~\bibnamefont
  {Hornberger}}, \bibinfo {author} {\bibfnamefont {P.~J.}\ \bibnamefont
  {Fagan}}, \bibinfo {author} {\bibfnamefont {J.}~\bibnamefont {T{\"u}xen}},
  \bibinfo {author} {\bibfnamefont {M.}~\bibnamefont {Mayor}},\ and\ \bibinfo
  {author} {\bibfnamefont {M.}~\bibnamefont {Arndt}},\ }\bibfield  {title}
  {\bibinfo {title} {Quantum interference of large organic molecules},\ }\href
  {https://doi.org/10.1038/ncomms1263} {\bibfield  {journal} {\bibinfo
  {journal} {Nature Communications}\ }\textbf {\bibinfo {volume} {2}},\
  \bibinfo {pages} {263} (\bibinfo {year} {2011})}\BibitemShut {NoStop}%
\bibitem [{\citenamefont {Eibenberger}\ \emph {et~al.}(2013)\citenamefont
  {Eibenberger}, \citenamefont {Gerlich}, \citenamefont {Arndt}, \citenamefont
  {Mayor},\ and\ \citenamefont {Tüxen}}]{Eibenberger_2013}%
  \BibitemOpen
  \bibfield  {author} {\bibinfo {author} {\bibfnamefont {S.}~\bibnamefont
  {Eibenberger}}, \bibinfo {author} {\bibfnamefont {S.}~\bibnamefont
  {Gerlich}}, \bibinfo {author} {\bibfnamefont {M.}~\bibnamefont {Arndt}},
  \bibinfo {author} {\bibfnamefont {M.}~\bibnamefont {Mayor}},\ and\ \bibinfo
  {author} {\bibfnamefont {J.}~\bibnamefont {Tüxen}},\ }\bibfield  {title}
  {\bibinfo {title} {Matter–wave interference of particles selected from a
  molecular library with masses exceeding 10 000 amu},\ }\href
  {https://doi.org/10.1039/C3CP51500A} {\bibfield  {journal} {\bibinfo
  {journal} {Phys. Chem. Chem. Phys.}\ }\textbf {\bibinfo {volume} {15}},\
  \bibinfo {pages} {14696} (\bibinfo {year} {2013})}\BibitemShut {NoStop}%
\bibitem [{\citenamefont {Romero-Isart}(2017)}]{Oriol_2017}%
  \BibitemOpen
  \bibfield  {author} {\bibinfo {author} {\bibfnamefont {O.}~\bibnamefont
  {Romero-Isart}},\ }\bibfield  {title} {\bibinfo {title} {Coherent inflation
  for large quantum superpositions of levitated microspheres},\ }\href
  {https://doi.org/10.1088/1367-2630/aa99bf} {\bibfield  {journal} {\bibinfo
  {journal} {New Journal of Physics}\ }\textbf {\bibinfo {volume} {19}},\
  \bibinfo {pages} {123029} (\bibinfo {year} {2017})},\ \Eprint
  {https://arxiv.org/abs/1612.04290} {arXiv:1612.04290 [quant-ph]} \BibitemShut
  {NoStop}%
\bibitem [{\citenamefont {Fein}\ \emph {et~al.}(2019)\citenamefont {Fein},
  \citenamefont {Geyer}, \citenamefont {Zwick}, \citenamefont {Kia{\l}ka},
  \citenamefont {Pedalino}, \citenamefont {Mayor}, \citenamefont {Gerlich},\
  and\ \citenamefont {Arndt}}]{Fein_2019}%
  \BibitemOpen
  \bibfield  {author} {\bibinfo {author} {\bibfnamefont {Y.~Y.}\ \bibnamefont
  {Fein}}, \bibinfo {author} {\bibfnamefont {P.}~\bibnamefont {Geyer}},
  \bibinfo {author} {\bibfnamefont {P.}~\bibnamefont {Zwick}}, \bibinfo
  {author} {\bibfnamefont {F.}~\bibnamefont {Kia{\l}ka}}, \bibinfo {author}
  {\bibfnamefont {S.}~\bibnamefont {Pedalino}}, \bibinfo {author}
  {\bibfnamefont {M.}~\bibnamefont {Mayor}}, \bibinfo {author} {\bibfnamefont
  {S.}~\bibnamefont {Gerlich}},\ and\ \bibinfo {author} {\bibfnamefont
  {M.}~\bibnamefont {Arndt}},\ }\bibfield  {title} {\bibinfo {title} {Quantum
  superposition of molecules beyond 25 kda},\ }\href
  {https://doi.org/10.1038/s41567-019-0663-9} {\bibfield  {journal} {\bibinfo
  {journal} {Nature Physics}\ }\textbf {\bibinfo {volume} {15}},\ \bibinfo
  {pages} {1242} (\bibinfo {year} {2019})}\BibitemShut {NoStop}%
\bibitem [{\citenamefont {Pino}\ \emph {et~al.}(2018)\citenamefont {Pino},
  \citenamefont {Prat-Camps}, \citenamefont {Sinha}, \citenamefont
  {Venkatesh},\ and\ \citenamefont {Romero-Isart}}]{Pino_2018}%
  \BibitemOpen
  \bibfield  {author} {\bibinfo {author} {\bibfnamefont {H.}~\bibnamefont
  {Pino}}, \bibinfo {author} {\bibfnamefont {J.}~\bibnamefont {Prat-Camps}},
  \bibinfo {author} {\bibfnamefont {K.}~\bibnamefont {Sinha}}, \bibinfo
  {author} {\bibfnamefont {B.~P.}\ \bibnamefont {Venkatesh}},\ and\ \bibinfo
  {author} {\bibfnamefont {O.}~\bibnamefont {Romero-Isart}},\ }\bibfield
  {title} {\bibinfo {title} {On-chip quantum interference of a superconducting
  microsphere},\ }\href {https://doi.org/10.1088/2058-9565/aa9d15} {\bibfield
  {journal} {\bibinfo  {journal} {Quantum Science and Technology}\ }\textbf
  {\bibinfo {volume} {3}},\ \bibinfo {pages} {025001} (\bibinfo {year}
  {2018})},\ \Eprint {https://arxiv.org/abs/1603.01553} {arXiv:1603.01553
  [quant-ph]} \BibitemShut {NoStop}%
\bibitem [{\citenamefont {Brand}\ \emph {et~al.}(2017)\citenamefont {Brand},
  \citenamefont {Eibenberger}, \citenamefont {Sezer},\ and\ \citenamefont
  {Arndt}}]{brand_2017}%
  \BibitemOpen
  \bibfield  {author} {\bibinfo {author} {\bibfnamefont {C.}~\bibnamefont
  {Brand}}, \bibinfo {author} {\bibfnamefont {S.}~\bibnamefont {Eibenberger}},
  \bibinfo {author} {\bibfnamefont {U.}~\bibnamefont {Sezer}},\ and\ \bibinfo
  {author} {\bibfnamefont {M.}~\bibnamefont {Arndt}},\ }\bibfield  {title}
  {\bibinfo {title} {Matter-wave physics with nanoparticles and biomolecules},\
  }\href@noop {} {\  (\bibinfo {year} {2017})},\ \Eprint
  {https://arxiv.org/abs/1703.02129} {arXiv:1703.02129 [quant-ph]} \BibitemShut
  {NoStop}%
\bibitem [{\citenamefont {Ford}(1982)}]{FORD_1982}%
  \BibitemOpen
  \bibfield  {author} {\bibinfo {author} {\bibfnamefont {L.~H.}\ \bibnamefont
  {Ford}},\ }\bibfield  {title} {\bibinfo {title} {Gravitational radiation by
  quantum systems},\ }\bibfield  {journal} {\bibinfo  {journal} {Ann. Phys.
  (N.Y.); (United States)}\ }\textbf {\bibinfo {volume} {144:2}},\ \href
  {https://doi.org/10.1016/0003-4916(82)90115-4} {10.1016/0003-4916(82)90115-4}
  (\bibinfo {year} {1982})\BibitemShut {NoStop}%
\bibitem [{\citenamefont {Lindner}\ and\ \citenamefont
  {Peres}(2005)}]{Lindner_2005}%
  \BibitemOpen
  \bibfield  {author} {\bibinfo {author} {\bibfnamefont {N.~H.}\ \bibnamefont
  {Lindner}}\ and\ \bibinfo {author} {\bibfnamefont {A.}~\bibnamefont
  {Peres}},\ }\bibfield  {title} {\bibinfo {title} {Testing quantum
  superpositions of the gravitational field with bose-einstein condensates},\
  }\href {https://doi.org/10.1103/PhysRevA.71.024101} {\bibfield  {journal}
  {\bibinfo  {journal} {Phys. Rev. A}\ }\textbf {\bibinfo {volume} {71}},\
  \bibinfo {pages} {024101} (\bibinfo {year} {2005})},\ \Eprint
  {https://arxiv.org/abs/0410030} {arXiv:0410030} \BibitemShut {NoStop}%
\bibitem [{\citenamefont {Bahrami}\ \emph {et~al.}(2015)\citenamefont
  {Bahrami}, \citenamefont {Bassi}, \citenamefont {McMillen}, \citenamefont
  {Paternostro},\ and\ \citenamefont {Ulbricht}}]{Bahrami_2015}%
  \BibitemOpen
  \bibfield  {author} {\bibinfo {author} {\bibfnamefont {M.}~\bibnamefont
  {Bahrami}}, \bibinfo {author} {\bibfnamefont {A.}~\bibnamefont {Bassi}},
  \bibinfo {author} {\bibfnamefont {S.}~\bibnamefont {McMillen}}, \bibinfo
  {author} {\bibfnamefont {M.}~\bibnamefont {Paternostro}},\ and\ \bibinfo
  {author} {\bibfnamefont {H.}~\bibnamefont {Ulbricht}},\ }\bibfield  {title}
  {\bibinfo {title} {Is gravity quantum?},\ }\href@noop {} {\  (\bibinfo {year}
  {2015})},\ \Eprint {https://arxiv.org/abs/1507.05733} {arXiv:1507.05733
  [quant-ph]} \BibitemShut {NoStop}%
\bibitem [{\citenamefont {Bose}\ \emph {et~al.}(2017)\citenamefont {Bose},
  \citenamefont {Mazumdar}, \citenamefont {Morley}, \citenamefont {Ulbricht},
  \citenamefont {Toro\ifmmode~\check{s}\else \v{s}\fi{}}, \citenamefont
  {Paternostro}, \citenamefont {Geraci}, \citenamefont {Barker}, \citenamefont
  {Kim},\ and\ \citenamefont {Milburn}}]{Bose_2017}%
  \BibitemOpen
  \bibfield  {author} {\bibinfo {author} {\bibfnamefont {S.}~\bibnamefont
  {Bose}}, \bibinfo {author} {\bibfnamefont {A.}~\bibnamefont {Mazumdar}},
  \bibinfo {author} {\bibfnamefont {G.~W.}\ \bibnamefont {Morley}}, \bibinfo
  {author} {\bibfnamefont {H.}~\bibnamefont {Ulbricht}}, \bibinfo {author}
  {\bibfnamefont {M.}~\bibnamefont {Toro\ifmmode~\check{s}\else \v{s}\fi{}}},
  \bibinfo {author} {\bibfnamefont {M.}~\bibnamefont {Paternostro}}, \bibinfo
  {author} {\bibfnamefont {A.~A.}\ \bibnamefont {Geraci}}, \bibinfo {author}
  {\bibfnamefont {P.~F.}\ \bibnamefont {Barker}}, \bibinfo {author}
  {\bibfnamefont {M.~S.}\ \bibnamefont {Kim}},\ and\ \bibinfo {author}
  {\bibfnamefont {G.}~\bibnamefont {Milburn}},\ }\bibfield  {title} {\bibinfo
  {title} {Spin entanglement witness for quantum gravity},\ }\href
  {https://doi.org/10.1103/PhysRevLett.119.240401} {\bibfield  {journal}
  {\bibinfo  {journal} {Phys. Rev. Lett.}\ }\textbf {\bibinfo {volume} {119}},\
  \bibinfo {pages} {240401} (\bibinfo {year} {2017})},\ \Eprint
  {https://arxiv.org/abs/1507.05733} {arXiv:1507.05733 [quant-ph]} \BibitemShut
  {NoStop}%
\bibitem [{\citenamefont {Marletto}\ and\ \citenamefont
  {Vedral}(2017)}]{Marletto_2017}%
  \BibitemOpen
  \bibfield  {author} {\bibinfo {author} {\bibfnamefont {C.}~\bibnamefont
  {Marletto}}\ and\ \bibinfo {author} {\bibfnamefont {V.}~\bibnamefont
  {Vedral}},\ }\bibfield  {title} {\bibinfo {title} {{Gravitationally-induced
  entanglement between two massive particles is sufficient evidence of quantum
  effects in gravity}},\ }\href
  {https://doi.org/10.1103/PhysRevLett.119.240402} {\bibfield  {journal}
  {\bibinfo  {journal} {Phys. Rev. Lett.}\ }\textbf {\bibinfo {volume} {119}},\
  \bibinfo {pages} {240402} (\bibinfo {year} {2017})},\ \Eprint
  {https://arxiv.org/abs/1707.06036} {arXiv:1707.06036 [quant-ph]} \BibitemShut
  {NoStop}%
\bibitem [{\citenamefont {Carney}\ \emph {et~al.}(2019)\citenamefont {Carney},
  \citenamefont {Stamp},\ and\ \citenamefont {Taylor}}]{Carney_2019}%
  \BibitemOpen
  \bibfield  {author} {\bibinfo {author} {\bibfnamefont {D.}~\bibnamefont
  {Carney}}, \bibinfo {author} {\bibfnamefont {P.~C.~E.}\ \bibnamefont
  {Stamp}},\ and\ \bibinfo {author} {\bibfnamefont {J.~M.}\ \bibnamefont
  {Taylor}},\ }\bibfield  {title} {\bibinfo {title} {Tabletop experiments for
  quantum gravity: a user\textquoteright{}s manual},\ }\href
  {https://doi.org/10.1088/1361-6382/aaf9ca} {\bibfield  {journal} {\bibinfo
  {journal} {Class. Quant. Grav.}\ }\textbf {\bibinfo {volume} {36}},\ \bibinfo
  {pages} {034001} (\bibinfo {year} {2019})},\ \Eprint
  {https://arxiv.org/abs/1807.11494} {arXiv:1807.11494 [quant-ph]} \BibitemShut
  {NoStop}%
\bibitem [{\citenamefont {Haine}(2021)}]{Haine_2021}%
  \BibitemOpen
  \bibfield  {author} {\bibinfo {author} {\bibfnamefont {S.~A.}\ \bibnamefont
  {Haine}},\ }\bibfield  {title} {\bibinfo {title} {Searching for signatures of
  quantum gravity in quantum gases},\ }\href
  {https://doi.org/10.1088/1367-2630/abd97d} {\bibfield  {journal} {\bibinfo
  {journal} {New J. Phys.}\ }\textbf {\bibinfo {volume} {23}},\ \bibinfo
  {pages} {033020} (\bibinfo {year} {2021})},\ \Eprint
  {https://arxiv.org/abs/1810.10202} {arXiv:1810.10202 [quant-ph]} \BibitemShut
  {NoStop}%
\bibitem [{\citenamefont {Qvarfort}\ \emph {et~al.}(2020)\citenamefont
  {Qvarfort}, \citenamefont {Bose},\ and\ \citenamefont
  {Serafini}}]{Qvarfort_2020}%
  \BibitemOpen
  \bibfield  {author} {\bibinfo {author} {\bibfnamefont {S.}~\bibnamefont
  {Qvarfort}}, \bibinfo {author} {\bibfnamefont {S.}~\bibnamefont {Bose}},\
  and\ \bibinfo {author} {\bibfnamefont {A.}~\bibnamefont {Serafini}},\
  }\bibfield  {title} {\bibinfo {title} {{Mesoscopic entanglement through
  central\textendash{}potential interactions}},\ }\href
  {https://doi.org/10.1088/1361-6455/abbe8d} {\bibfield  {journal} {\bibinfo
  {journal} {J. Phys. B}\ }\textbf {\bibinfo {volume} {53}},\ \bibinfo {pages}
  {235501} (\bibinfo {year} {2020})},\ \Eprint
  {https://arxiv.org/abs/1812.09776} {arXiv:1812.09776 [quant-ph]} \BibitemShut
  {NoStop}%
\bibitem [{\citenamefont {Carlesso}\ \emph {et~al.}(2019)\citenamefont
  {Carlesso}, \citenamefont {Bassi}, \citenamefont {Paternostro},\ and\
  \citenamefont {Ulbricht}}]{Carlesso_2019}%
  \BibitemOpen
  \bibfield  {author} {\bibinfo {author} {\bibfnamefont {M.}~\bibnamefont
  {Carlesso}}, \bibinfo {author} {\bibfnamefont {A.}~\bibnamefont {Bassi}},
  \bibinfo {author} {\bibfnamefont {M.}~\bibnamefont {Paternostro}},\ and\
  \bibinfo {author} {\bibfnamefont {H.}~\bibnamefont {Ulbricht}},\ }\bibfield
  {title} {\bibinfo {title} {Testing the gravitational field generated by a
  quantum superposition},\ }\href {https://doi.org/10.1088/1367-2630/ab41c1}
  {\bibfield  {journal} {\bibinfo  {journal} {New J. Phys.}\ }\textbf {\bibinfo
  {volume} {21}},\ \bibinfo {pages} {093052} (\bibinfo {year} {2019})},\
  \Eprint {https://arxiv.org/abs/1906.04513} {arXiv:1906.04513 [quant-ph]}
  \BibitemShut {NoStop}%
\bibitem [{\citenamefont {Howl}\ \emph {et~al.}(2021)\citenamefont {Howl},
  \citenamefont {Vedral}, \citenamefont {Naik}, \citenamefont {Christodoulou},
  \citenamefont {Rovelli},\ and\ \citenamefont {Iyer}}]{Howl_2021}%
  \BibitemOpen
  \bibfield  {author} {\bibinfo {author} {\bibfnamefont {R.}~\bibnamefont
  {Howl}}, \bibinfo {author} {\bibfnamefont {V.}~\bibnamefont {Vedral}},
  \bibinfo {author} {\bibfnamefont {D.}~\bibnamefont {Naik}}, \bibinfo {author}
  {\bibfnamefont {M.}~\bibnamefont {Christodoulou}}, \bibinfo {author}
  {\bibfnamefont {C.}~\bibnamefont {Rovelli}},\ and\ \bibinfo {author}
  {\bibfnamefont {A.}~\bibnamefont {Iyer}},\ }\bibfield  {title} {\bibinfo
  {title} {Non-gaussianity as a signature of a quantum theory of gravity},\
  }\href {https://doi.org/10.1103/PRXQuantum.2.010325} {\bibfield  {journal}
  {\bibinfo  {journal} {P.R.X. Quantum.}\ }\textbf {\bibinfo {volume} {2}},\
  \bibinfo {pages} {010325} (\bibinfo {year} {2021})},\ \Eprint
  {https://arxiv.org/abs/2004.01189} {arXiv:2004.01189 [quant-ph]} \BibitemShut
  {NoStop}%
\bibitem [{\citenamefont {Matsumura}\ and\ \citenamefont
  {Yamamoto}(2020)}]{Matsumura_2020}%
  \BibitemOpen
  \bibfield  {author} {\bibinfo {author} {\bibfnamefont {A.}~\bibnamefont
  {Matsumura}}\ and\ \bibinfo {author} {\bibfnamefont {K.}~\bibnamefont
  {Yamamoto}},\ }\bibfield  {title} {\bibinfo {title} {Gravity-induced
  entanglement in optomechanical systems},\ }\href
  {https://doi.org/10.1103/PhysRevD.102.106021} {\bibfield  {journal} {\bibinfo
   {journal} {Phys. Rev. D}\ }\textbf {\bibinfo {volume} {102}},\ \bibinfo
  {pages} {106021} (\bibinfo {year} {2020})},\ \Eprint
  {https://arxiv.org/abs/2010.05161} {arXiv:2010.05161 [gr-qc]} \BibitemShut
  {NoStop}%
\bibitem [{\citenamefont {Pedernales}\ \emph {et~al.}(2021)\citenamefont
  {Pedernales}, \citenamefont {Streltsov},\ and\ \citenamefont
  {Plenio}}]{Pedernales_2021}%
  \BibitemOpen
  \bibfield  {author} {\bibinfo {author} {\bibfnamefont {J.~S.}\ \bibnamefont
  {Pedernales}}, \bibinfo {author} {\bibfnamefont {K.}~\bibnamefont
  {Streltsov}},\ and\ \bibinfo {author} {\bibfnamefont {M.~B.}\ \bibnamefont
  {Plenio}},\ }\bibfield  {title} {\bibinfo {title} {Enhancing gravitational
  interaction between quantum systems by a massive mediator},\ }\href@noop {}
  {\  (\bibinfo {year} {2021})},\ \Eprint {https://arxiv.org/abs/2104.14524}
  {arXiv:2104.14524 [quant-ph]} \BibitemShut {NoStop}%
\bibitem [{\citenamefont {Liu}\ \emph {et~al.}(2021)\citenamefont {Liu},
  \citenamefont {Mummery}, \citenamefont {Zhou},\ and\ \citenamefont
  {Sillanp\"a\"a}}]{Liu_2021}%
  \BibitemOpen
  \bibfield  {author} {\bibinfo {author} {\bibfnamefont {Y.}~\bibnamefont
  {Liu}}, \bibinfo {author} {\bibfnamefont {J.}~\bibnamefont {Mummery}},
  \bibinfo {author} {\bibfnamefont {J.}~\bibnamefont {Zhou}},\ and\ \bibinfo
  {author} {\bibfnamefont {M.~A.}\ \bibnamefont {Sillanp\"a\"a}},\ }\bibfield
  {title} {\bibinfo {title} {Gravitational forces between nonclassical
  mechanical oscillators},\ }\href
  {https://doi.org/10.1103/PhysRevApplied.15.034004} {\bibfield  {journal}
  {\bibinfo  {journal} {Phys. Rev. Applied}\ }\textbf {\bibinfo {volume}
  {15}},\ \bibinfo {pages} {034004} (\bibinfo {year} {2021})}\BibitemShut
  {NoStop}%
\bibitem [{\citenamefont {Datta}\ and\ \citenamefont
  {Miao}(2021)}]{Datta_2021}%
  \BibitemOpen
  \bibfield  {author} {\bibinfo {author} {\bibfnamefont {A.}~\bibnamefont
  {Datta}}\ and\ \bibinfo {author} {\bibfnamefont {H.}~\bibnamefont {Miao}},\
  }\bibfield  {title} {\bibinfo {title} {{Signatures of the quantum nature of
  gravity in the differential motion of two masses}},\ }\href
  {https://doi.org/10.1088/2058-9565/ac1adf} {\bibfield  {journal} {\bibinfo
  {journal} {Quantum Sci. Technol.}\ }\textbf {\bibinfo {volume} {6}},\
  \bibinfo {pages} {045014} (\bibinfo {year} {2021})},\ \Eprint
  {https://arxiv.org/abs/2104.04414} {arXiv:2104.04414 [gr-qc]} \BibitemShut
  {NoStop}%
\bibitem [{\citenamefont {Gonzalez-Ballestero}\ \emph
  {et~al.}(2021)\citenamefont {Gonzalez-Ballestero}, \citenamefont
  {Aspelmeyer}, \citenamefont {Novotny}, \citenamefont {Quidant},\ and\
  \citenamefont {Romero-Isart}}]{Gonzalez-Ballestero:2021}%
  \BibitemOpen
  \bibfield  {author} {\bibinfo {author} {\bibfnamefont {C.}~\bibnamefont
  {Gonzalez-Ballestero}}, \bibinfo {author} {\bibfnamefont {M.}~\bibnamefont
  {Aspelmeyer}}, \bibinfo {author} {\bibfnamefont {L.}~\bibnamefont {Novotny}},
  \bibinfo {author} {\bibfnamefont {R.}~\bibnamefont {Quidant}},\ and\ \bibinfo
  {author} {\bibfnamefont {O.}~\bibnamefont {Romero-Isart}},\ }\bibfield
  {title} {\bibinfo {title} {Levitodynamics: Levitation and control of
  microscopic objects in vacuum},\ }\href
  {https://doi.org/10.1126/science.abg3027} {\bibfield  {journal} {\bibinfo
  {journal} {Science}\ }\textbf {\bibinfo {volume} {374}},\ \bibinfo {pages}
  {3027} (\bibinfo {year} {2021})},\ \Eprint {https://arxiv.org/abs/2111.05215}
  {arXiv:2111.05215 [quant-ph]} \BibitemShut {NoStop}%
\bibitem [{\citenamefont {Krisnanda}\ \emph {et~al.}(2020)\citenamefont
  {Krisnanda}, \citenamefont {Tham}, \citenamefont {Paternostro},\ and\
  \citenamefont {Paterek}}]{Krisnanda:2020uh}%
  \BibitemOpen
  \bibfield  {author} {\bibinfo {author} {\bibfnamefont {T.}~\bibnamefont
  {Krisnanda}}, \bibinfo {author} {\bibfnamefont {G.~Y.}\ \bibnamefont {Tham}},
  \bibinfo {author} {\bibfnamefont {M.}~\bibnamefont {Paternostro}},\ and\
  \bibinfo {author} {\bibfnamefont {T.}~\bibnamefont {Paterek}},\ }\bibfield
  {title} {\bibinfo {title} {Observable quantum entanglement due to gravity},\
  }\href {https://doi.org/10.1038/s41534-020-0243-y} {\bibfield  {journal}
  {\bibinfo  {journal} {npj Quantum Information}\ }\textbf {\bibinfo {volume}
  {6}},\ \bibinfo {pages} {12} (\bibinfo {year} {2020})}\BibitemShut {NoStop}%
\bibitem [{\citenamefont {Margalit}\ \emph {et~al.}(2021)\citenamefont
  {Margalit}, \citenamefont {Dobkowski}, \citenamefont {Zhou}, \citenamefont
  {Amit}, \citenamefont {Japha}, \citenamefont {Moukouri}, \citenamefont
  {Rohrlich}, \citenamefont {Mazumdar}, \citenamefont {Bose}, \citenamefont
  {Henkel},\ and\ \citenamefont {Folman}}]{doi:10.1126/sciadv.abg2879}%
  \BibitemOpen
  \bibfield  {author} {\bibinfo {author} {\bibfnamefont {Y.}~\bibnamefont
  {Margalit}}, \bibinfo {author} {\bibfnamefont {O.}~\bibnamefont {Dobkowski}},
  \bibinfo {author} {\bibfnamefont {Z.}~\bibnamefont {Zhou}}, \bibinfo {author}
  {\bibfnamefont {O.}~\bibnamefont {Amit}}, \bibinfo {author} {\bibfnamefont
  {Y.}~\bibnamefont {Japha}}, \bibinfo {author} {\bibfnamefont
  {S.}~\bibnamefont {Moukouri}}, \bibinfo {author} {\bibfnamefont
  {D.}~\bibnamefont {Rohrlich}}, \bibinfo {author} {\bibfnamefont
  {A.}~\bibnamefont {Mazumdar}}, \bibinfo {author} {\bibfnamefont
  {S.}~\bibnamefont {Bose}}, \bibinfo {author} {\bibfnamefont {C.}~\bibnamefont
  {Henkel}},\ and\ \bibinfo {author} {\bibfnamefont {R.}~\bibnamefont
  {Folman}},\ }\bibfield  {title} {\bibinfo {title} {Realization of a complete
  stern-gerlach interferometer: Toward a test of quantum gravity},\ }\href
  {https://doi.org/10.1126/sciadv.abg2879} {\bibfield  {journal} {\bibinfo
  {journal} {Science Advances}\ }\textbf {\bibinfo {volume} {7}},\ \bibinfo
  {pages} {eabg2879} (\bibinfo {year} {2021})},\ \Eprint
  {https://arxiv.org/abs/https://www.science.org/doi/pdf/10.1126/sciadv.abg2879}
  {https://www.science.org/doi/pdf/10.1126/sciadv.abg2879} \BibitemShut
  {NoStop}%
\bibitem [{\citenamefont {Christodoulou}\ and\ \citenamefont
  {Rovelli}(2019)}]{CHRISTODOULOU201964}%
  \BibitemOpen
  \bibfield  {author} {\bibinfo {author} {\bibfnamefont {M.}~\bibnamefont
  {Christodoulou}}\ and\ \bibinfo {author} {\bibfnamefont {C.}~\bibnamefont
  {Rovelli}},\ }\bibfield  {title} {\bibinfo {title} {On the possibility of
  laboratory evidence for quantum superposition of geometries},\ }\href
  {https://doi.org/https://doi.org/10.1016/j.physletb.2019.03.015} {\bibfield
  {journal} {\bibinfo  {journal} {Physics Letters B}\ }\textbf {\bibinfo
  {volume} {792}},\ \bibinfo {pages} {64} (\bibinfo {year} {2019})}\BibitemShut
  {NoStop}%
\bibitem [{\citenamefont {Bose}\ \emph {et~al.}(2022)\citenamefont {Bose},
  \citenamefont {Mazumdar}, \citenamefont {Schut},\ and\ \citenamefont
  {Toro\v{s}}}]{Bose:2022uxe}%
  \BibitemOpen
  \bibfield  {author} {\bibinfo {author} {\bibfnamefont {S.}~\bibnamefont
  {Bose}}, \bibinfo {author} {\bibfnamefont {A.}~\bibnamefont {Mazumdar}},
  \bibinfo {author} {\bibfnamefont {M.}~\bibnamefont {Schut}},\ and\ \bibinfo
  {author} {\bibfnamefont {M.}~\bibnamefont {Toro\v{s}}},\ }\bibfield  {title}
  {\bibinfo {title} {{Mechanism for the quantum natured gravitons to entangle
  masses}},\ }\href@noop {} {\  (\bibinfo {year} {2022})},\ \Eprint
  {https://arxiv.org/abs/2201.03583} {arXiv:2201.03583 [gr-qc]} \BibitemShut
  {NoStop}%
\bibitem [{\citenamefont {Mari}\ \emph {et~al.}(2016)\citenamefont {Mari},
  \citenamefont {De~Palma},\ and\ \citenamefont {Giovannetti}}]{Mari2009}%
  \BibitemOpen
  \bibfield  {author} {\bibinfo {author} {\bibfnamefont {A.}~\bibnamefont
  {Mari}}, \bibinfo {author} {\bibfnamefont {G.}~\bibnamefont {De~Palma}},\
  and\ \bibinfo {author} {\bibfnamefont {V.}~\bibnamefont {Giovannetti}},\
  }\bibfield  {title} {\bibinfo {title} {Experiments testing macroscopic
  quantum superpositions must be slow},\ }\href
  {https://doi.org/10.1038/srep22777} {\bibfield  {journal} {\bibinfo
  {journal} {Scientific Reports}\ }\textbf {\bibinfo {volume} {6}},\ \bibinfo
  {pages} {22777} (\bibinfo {year} {2016})},\ \Eprint
  {https://arxiv.org/abs/1509.02408} {arXiv:1509.02408 [quant-ph]} \BibitemShut
  {NoStop}%
\bibitem [{\citenamefont {Anastopoulos}\ and\ \citenamefont
  {Hu}(2018)}]{Anastopoulos_2018}%
  \BibitemOpen
  \bibfield  {author} {\bibinfo {author} {\bibfnamefont {C.}~\bibnamefont
  {Anastopoulos}}\ and\ \bibinfo {author} {\bibfnamefont {B.-L.}\ \bibnamefont
  {Hu}},\ }\bibfield  {title} {\bibinfo {title} {Comment on ``a spin
  entanglement witness for quantum gravity'' and on ``gravitationally induced
  entanglement between two massive particles is sufficient evidence of quantum
  effects in gravity''},\ }\href@noop {} {\  (\bibinfo {year} {2018})},\
  \Eprint {https://arxiv.org/abs/1804.11315} {arXiv:1804.11315 [quant-ph]}
  \BibitemShut {NoStop}%
\bibitem [{\citenamefont {Carney}(2022)}]{Carney_2021}%
  \BibitemOpen
  \bibfield  {author} {\bibinfo {author} {\bibfnamefont {D.}~\bibnamefont
  {Carney}},\ }\bibfield  {title} {\bibinfo {title} {Newton, entanglement, and
  the graviton},\ }\href {https://doi.org/10.1103/PhysRevD.105.024029}
  {\bibfield  {journal} {\bibinfo  {journal} {Phys. Rev. D}\ }\textbf {\bibinfo
  {volume} {105}},\ \bibinfo {pages} {024029} (\bibinfo {year}
  {2022})}\BibitemShut {NoStop}%
\bibitem [{\citenamefont {Rydving}\ \emph {et~al.}(2021)\citenamefont
  {Rydving}, \citenamefont {Aurell},\ and\ \citenamefont
  {Pikovski}}]{Rydving_2021}%
  \BibitemOpen
  \bibfield  {author} {\bibinfo {author} {\bibfnamefont {E.}~\bibnamefont
  {Rydving}}, \bibinfo {author} {\bibfnamefont {E.}~\bibnamefont {Aurell}},\
  and\ \bibinfo {author} {\bibfnamefont {I.}~\bibnamefont {Pikovski}},\
  }\bibfield  {title} {\bibinfo {title} {{Do Gedankenexperiments compel
  quantization of gravity?}},\ }\href
  {https://doi.org/10.1103/PhysRevD.104.086024} {\bibfield  {journal} {\bibinfo
   {journal} {Phys. Rev. D}\ }\textbf {\bibinfo {volume} {104}},\ \bibinfo
  {pages} {086024} (\bibinfo {year} {2021})},\ \Eprint
  {https://arxiv.org/abs/2107.07514} {arXiv:2107.07514 [gr-qc]} \BibitemShut
  {NoStop}%
\bibitem [{\citenamefont {Baym}\ and\ \citenamefont {Ozawa}(2009)}]{Baym_2009}%
  \BibitemOpen
  \bibfield  {author} {\bibinfo {author} {\bibfnamefont {G.}~\bibnamefont
  {Baym}}\ and\ \bibinfo {author} {\bibfnamefont {T.}~\bibnamefont {Ozawa}},\
  }\bibfield  {title} {\bibinfo {title} {Two-slit diffraction with highly
  charged particles: Niels bohr's consistency argument that the electromagnetic
  field must be quantized},\ }\href {https://doi.org/10.1073/pnas.0813239106}
  {\bibfield  {journal} {\bibinfo  {journal} {Proc. Nat. Acad. Sci.}\ }\textbf
  {\bibinfo {volume} {106}},\ \bibinfo {pages} {3035} (\bibinfo {year}
  {2009})},\ \Eprint {https://arxiv.org/abs/0902.2615} {arXiv:0902.2615
  [quant-ph]} \BibitemShut {NoStop}%
\bibitem [{\citenamefont {Unruh}(2000)}]{Unruh_2000}%
  \BibitemOpen
  \bibfield  {author} {\bibinfo {author} {\bibfnamefont {W.~G.}\ \bibnamefont
  {Unruh}},\ }\bibfield  {title} {\bibinfo {title} {False loss of coherence},\
  }in\ \href@noop {} {\emph {\bibinfo {booktitle} {Relativistic Quantum
  Measurement and Decoherence}}},\ \bibinfo {editor} {edited by\ \bibinfo
  {editor} {\bibfnamefont {H.-P.}\ \bibnamefont {Breuer}}\ and\ \bibinfo
  {editor} {\bibfnamefont {F.}~\bibnamefont {Petruccione}}}\ (\bibinfo
  {publisher} {Springer Berlin Heidelberg},\ \bibinfo {address} {Berlin,
  Heidelberg},\ \bibinfo {year} {2000})\ pp.\ \bibinfo {pages}
  {125--140}\BibitemShut {NoStop}%
\bibitem [{\citenamefont {Prabhu}\ \emph {et~al.}(2022)\citenamefont {Prabhu},
  \citenamefont {Satishchandran},\ and\ \citenamefont
  {Wald}}]{Satishchandran_2022}%
  \BibitemOpen
  \bibfield  {author} {\bibinfo {author} {\bibfnamefont {K.}~\bibnamefont
  {Prabhu}}, \bibinfo {author} {\bibfnamefont {G.}~\bibnamefont
  {Satishchandran}},\ and\ \bibinfo {author} {\bibfnamefont {R.~M.}\
  \bibnamefont {Wald}},\ }\bibfield  {title} {\bibinfo {title} {{Infrared
  finite scattering theory in quantum field theory and quantum gravity}},\
  }\href {https://doi.org/10.1103/PhysRevD.106.066005} {\bibfield  {journal}
  {\bibinfo  {journal} {Phys. Rev. D}\ }\textbf {\bibinfo {volume} {106}},\
  \bibinfo {pages} {066005} (\bibinfo {year} {2022})},\ \Eprint
  {https://arxiv.org/abs/2203.14334} {arXiv:2203.14334 [hep-th]} \BibitemShut
  {NoStop}%
\bibitem [{\citenamefont {Kulish}\ and\ \citenamefont
  {Faddeev}(1970)}]{Kulish_1970}%
  \BibitemOpen
  \bibfield  {author} {\bibinfo {author} {\bibfnamefont {P.~P.}\ \bibnamefont
  {Kulish}}\ and\ \bibinfo {author} {\bibfnamefont {L.~D.}\ \bibnamefont
  {Faddeev}},\ }\bibfield  {title} {\bibinfo {title} {Asymptotic conditions and
  infrared divergences in quantum electrodynamics},\ }\href
  {https://doi.org/10.1007/BF01066485} {\bibfield  {journal} {\bibinfo
  {journal} {Theor. Math. Phys.}\ }\textbf {\bibinfo {volume} {4}},\ \bibinfo
  {pages} {745} (\bibinfo {year} {1970})}\BibitemShut {NoStop}%
\bibitem [{\citenamefont {Bloch}\ and\ \citenamefont
  {Nordsieck}(1937)}]{Bloch_1937}%
  \BibitemOpen
  \bibfield  {author} {\bibinfo {author} {\bibfnamefont {F.}~\bibnamefont
  {Bloch}}\ and\ \bibinfo {author} {\bibfnamefont {A.}~\bibnamefont
  {Nordsieck}},\ }\bibfield  {title} {\bibinfo {title} {Note on the radiation
  field of the electron},\ }\href {https://doi.org/10.1103/PhysRev.52.54}
  {\bibfield  {journal} {\bibinfo  {journal} {Phys. Rev.}\ }\textbf {\bibinfo
  {volume} {52}},\ \bibinfo {pages} {54} (\bibinfo {year} {1937})}\BibitemShut
  {NoStop}%
\bibitem [{\citenamefont {Unruh}\ and\ \citenamefont
  {Wald}(1984)}]{Unruh_1984}%
  \BibitemOpen
  \bibfield  {author} {\bibinfo {author} {\bibfnamefont {W.~G.}\ \bibnamefont
  {Unruh}}\ and\ \bibinfo {author} {\bibfnamefont {R.~M.}\ \bibnamefont
  {Wald}},\ }\bibfield  {title} {\bibinfo {title} {What happens when an
  accelerating observer detects a rindler particle},\ }\href
  {https://doi.org/10.1103/PhysRevD.29.1047} {\bibfield  {journal} {\bibinfo
  {journal} {Phys. Rev. D}\ }\textbf {\bibinfo {volume} {29}},\ \bibinfo
  {pages} {1047} (\bibinfo {year} {1984})}\BibitemShut {NoStop}%
\end{thebibliography}%

\end{document}